\documentclass[structabstract]{aa}  
\usepackage{graphicx}
\usepackage{longtable}
\usepackage{appendix}

\usepackage{txfonts}
\graphicspath {{espectros/}{/}{reducedspectra00}}
\begin{document}
   \title{The chemical footprint of the star formation feedback in M~82 on scales of $\sim$100~pc\footnotemark }


   \author{D. Ginard\inst{1}
          \and
          A. Fuente\inst{1}
          \and
          S. Garc\'ia-Burillo\inst{2}
          \and
          T. Alonso-Albi\inst{2}
                    \and
          M. Krips\inst{3}
          \and
          M. Gerin\inst{4}
          \and
          R. Neri\inst{3}
          \and
          P. Pilleri\inst{5,6}
          \and
          A. Usero\inst{2}
          \and
          S.P. Trevi\~no-Morales\inst{7}
          }
\institute{ Observatorio Astron\'omico Nacional (OAN, IGN), Apdo. 112,
             E-28800 Alcal\'a de Henares, Madrid, Spain
         \and
         Observatorio Astron\'omico Nacional (OAN, IGN), Observatorio de Madrid, Alfonso XII, 3, E-28014, Madrid
         \and
            Institut de Radioastronomie Millim\'etrique, 300 rue de la Piscine, F-38406 St Martin d'H\`eres, France
          \and
           CNRS UMR8112, LERMA, Observatoire de Paris and Ecole Normale Sup\'erieure. 24 Rue Lhomond, 75231 Paris cedex 05, France
          \and
           Universit\'e de Toulouse; UPS-OMP; IRAP;  Toulouse, France
           \and
           CNRS; IRAP; 9 Av. colonel Roche, BP 44346, F-31028 Toulouse cedex 4, France 
          \and          
           Instituto de Radioastronom\'{\i}a Milim\'etrica (IRAM-Spain), Ave. Divina Pastora, 7, E-18012 Granada, Spain   }
   \date{Received September , 2014; accepted December --, 2014}

 
  \abstract
   {M~82 is one of the nearest and brightest starburst galaxies. It has been extensively studied in the last decade and
by now it is considered as the prototypical extragalactic PDR and a reference for the study of the star formation feedback. }
   {Our aim is to characterize the molecular chemistry in M~82 at spatial scales of giant molecular clouds (GMCs), $\sim$100~pc, to 
investigate the feedback effects of the star formation activity.}
   {We present interferometric observations
of the CN 1$\rightarrow$0 (113.491~GHz), 
N$_2$H$^+$ 1$\rightarrow$0 (93.173~GHz), H(41)$\alpha$ (92.034~GHz), CH$_3$CN (91.987~GHz), CS 3$\rightarrow$2 (146.969~GHz), 
c-C$_3$H$_2$ 3$_{1,2}$$\rightarrow$2$_{2,1}$ (145.089~GHz), H$_2$CO 2$_{0,2}$$\rightarrow$1$_{0,1}$ (145.603~GHz) and 
HC$_3$N 16$\rightarrow$15 (145.601~GHz) lines carried out with the IRAM Plateau de Bure Interferometer (PdBI).
PDR chemical modelling is used to interpret these observations.}
   {Our results show that the abundances of N$_2$H$^+$, CS and H$^{13}$CO$^+$ remain quite constant across the galaxy confirming that these species
are excellent tracers of the dense molecular gas. On the contrary, the abundance of CN increases by a factor of $\sim$3 in the inner $x2$ bar orbits.
The [CN]/[N$_2$H$^+$] ratio is well correlated with the H(41)$\alpha$ emission at all spatial scales down to $\sim$100 pc. Chemical modelling shows that
the variations in the [CN]/[N$_2$H$^+$] ratio can be explained as the consequence of differences in the local intestellar UV field and in the average cloud sizes 
within the nucleus of the galaxy. }
   {Our high-spatial resolution imaging of the starburst galaxy M~82 shows that the star formation activity has a strong impact on the chemistry of the molecular
gas. In particular, the entire nucleus behaves as a giant photon-dominated region (PDR) whose chemistry is determined by the local UV flux. The detection
of N$_2$H$^+$ shows the existence of a population of clouds with A$_v$$>$20~mag all across the galaxy plane.
These clouds constitute the molecular gas reservoir for the formation of new stars
and, although distributed all along the nucleus, the highest concentration occurs in the outer $x1$ bar orbits (R$\sim$280~pc). }

   \maketitle
\footnotetext{Based on observations carried out with the IRAM Plateau de Bure Interferometer. IRAM is supported by INSU/CNRS (France), MPG (Germany), and IGN (Spain).}

\section{Introduction}

M~82 is one of the nearest and brightest starburst galaxies. Located at a distance of 3.9~Mpc, 
and with a luminosity 
of 3.7$\times$10$^{10}$~L$_{\odot}$, it has been the subject of 
continuum and line observations at all wavelengths from
X-rays to the radio domains.
Several molecular line studies indicate that the strong UV
field has heavily influenced the physical conditions, kinematics and chemistry of the interstellar gas
(Mao et al. 2000; Weiss et al. 2001a,b; Garc\'{\i}a-Burillo et al. 2001, 2002; Fuente et al. 2005, 2006, 2008; Aladro et al. 2011). 
In fact M~82 has become the prototypical 
extragalactic PDR and a reference for the study and interpretation of star formation feedback in extreme starbursts near and far.

\begin{table*}
\caption{Observations}             
\begin{center}    
\begin{tabular}{llrllcl} \\  \hline
\multicolumn{2}{c}{Line} & \multicolumn{1}{c}{Freq.} & \multicolumn{1}{c}{beam} & \multicolumn{1}{c}{Fields} &
\multicolumn{1}{c}{T$_b$/Flux} & \multicolumn{1}{c}{Date or ref$^1$} \\
\multicolumn{2}{c}{} & \multicolumn{1}{l}{(GHz)} & \multicolumn{1}{c}{($\arcsec$)} & \multicolumn{1}{c}{} 
& \multicolumn{1}{c}{K/(Jy/beam)} & \multicolumn{1}{c}{} \\
\hline
C$^{18}$O   & 1$\rightarrow$0     & 109.782     & 3.8$\arcsec$$\times$3.5$\arcsec$ PA -4$^\circ$  &  (0$\arcsec$, 0$\arcsec$) & 
7.6 & W01 (Zero-spacing) \\
CO  & 2$\rightarrow$1 & 230.538    & 1.5$\arcsec$$\times$1.4$\arcsec$  PA 0$^\circ$    & 
(-30$\arcsec$, -8$\arcsec$), (-20$\arcsec$, -6$\arcsec$) & 11  &  W01 (Zero-spacing) \\
 & &    &    & (-10$\arcsec$, -4$\arcsec$), (10$\arcsec$, 4$\arcsec$) &   &       \\
 & &    &    & (20$\arcsec$, 8$\arcsec$), (30$\arcsec$, 10$\arcsec$)         &   &      \\
CN & 1$\rightarrow$0 &  113.491   & 2.5$\arcsec$$\times$2.3$\arcsec$ PA 170$^\circ$ & (0$\arcsec$, 0$\arcsec$)  & 17  &  May-Dec 2005 \\ 
H$^{13}$CO$^+$ &  1$\rightarrow$0 & 86.754 & 5.9$\arcsec$$\times$5.6$\arcsec$ PA 105$^\circ$ & (0$\arcsec$, 0$\arcsec$) & 5   &  G01 \\ 
HCO            &  1$\rightarrow$0 & 86.670 & 5.9$\arcsec$$\times$5.6$\arcsec$ PA 105$^\circ$ & (0$\arcsec$, 0$\arcsec$) & 5   &  G01 \\ 
HOC$^+$        &  1$\rightarrow$0 & 89.487 & 4.4$\arcsec$$\times$3.6$\arcsec$ PA 128$^\circ$ & (0$\arcsec$, 0$\arcsec$) & 9   &  F08  \\   
N$_2$H$^+$     &  1$\rightarrow$0 & 93.173 & 3.6$\arcsec$$\times$2.7$\arcsec$ PA 94$^\circ$  & (0$\arcsec$, 0$\arcsec$) & 14  & Dec 2010 - May 2011 \\  
H    & (41) $\alpha$     & 92.034     & 3.5$\arcsec$$\times$2.8$\arcsec$ PA 94$^\circ$   &  (0$\arcsec$, 0$\arcsec$) & 15 & Dec 2010 - May 2011 \\
CH$_3$CN       & 5$_3$$\rightarrow$4$_0$ & 91.971 & 3.5$\arcsec$$\times$2.8$\arcsec$ PA 94$^\circ$ &  (0$\arcsec$, 0$\arcsec$) & 15 & Dec 2010 - May 2011 \\
               & 5$_2$$\rightarrow$4$_2$ & 91.980 & 3.5$\arcsec$$\times$2.8$\arcsec$ PA 94$^\circ$ &  (0$\arcsec$, 0$\arcsec$) & 15 & Dec 2010 - May 2011 \\
               & 5$_1$$\rightarrow$4$_1$ & 91.985 & 3.5$\arcsec$$\times$2.8$\arcsec$ PA 94$^\circ$ &  (0$\arcsec$, 0$\arcsec$) & 15 & Dec 2010 - May 2011 \\
               & 5$_0$$\rightarrow$4$_0$ & 91.987 & 3.5$\arcsec$$\times$2.8$\arcsec$ PA 94$^\circ$ &  (0$\arcsec$, 0$\arcsec$) & 15 & Dec 2010 - May 2011 \\
H              &  (35) $\alpha$      & 147.047  & 2.3$\arcsec$$\times$1.9$\arcsec$ PA 75$^\circ$ & (+7$\arcsec$, 
+2.5$\arcsec$), (-7$\arcsec$, -2.5$\arcsec$) & 13 & Dec 2010 - May 2011 \\
CS             &  3$\rightarrow$2 & 146.969  & 2.3$\arcsec$$\times$1.9$\arcsec$ PA 75$^\circ$ & (+7$\arcsec$, 
+2.5$\arcsec$), (-7$\arcsec$, -2.5$\arcsec$) & 12 & Dec 2010 - May 2011 \\
c-C$_3$H$_2$   &  3$_{1.2}$$\rightarrow$2$_{2,1}$ & 145.089  & 2.3$\arcsec$$\times$1.9$\arcsec$ PA 75$^\circ$ & (+7$\arcsec$, 
+2.5$\arcsec$), (-7$\arcsec$, -2.5$\arcsec$) & 12 & Dec 2010 - May 2011 \\
A-CH$_3$OH     & 3$_{0,0}$$\rightarrow$2$_{0,0}$  & 145.103  &2.3$\arcsec$$\times$1.9$\arcsec$ PA 75$^\circ$ & (+7$\arcsec$, 
+2.5$\arcsec$), (-7$\arcsec$, -2.5$\arcsec$) & 12 & Dec 2010 - May 2011 \\
E-CH$_3$OH     & 3$_{0,0}$$\rightarrow$2$_{0,0}$  & 145.094  &2.3$\arcsec$$\times$1.9$\arcsec$ PA 75$^\circ$ & (+7$\arcsec$, 
+2.5$\arcsec$), (-7$\arcsec$, -2.5$\arcsec$) & 12 & Dec 2010 - May 2011 \\
               & 3$_{-1,0}$$\rightarrow$2$_{-1,0}$ & 145.097 &2.3$\arcsec$$\times$1.9$\arcsec$ PA 75$^\circ$ & (+7$\arcsec$, 
+2.5$\arcsec$), (-7$\arcsec$, -2.5$\arcsec$) & 12  & Dec 2010 - May 2011 \\

H$_2$CO        &  2$_{0,2}$$\rightarrow$1$_{0,1}$ & 145.603 & 2.3$\arcsec$$\times$1.9$\arcsec$ PA 75$^\circ$ & 
(+7$\arcsec$, +2.5$\arcsec$), (-7$\arcsec$, -2.5$\arcsec$) & 12  & Dec 2010 - May 2011 \\
HC$_3$N        &  16$\rightarrow$15               & 145.561 & 2.3$\arcsec$$\times$1.9$\arcsec$ PA 75$^\circ$ & 
(+7$\arcsec$, +2.5$\arcsec$), (-7$\arcsec$, -2.5$\arcsec$)  & 12 & Dec 2010 - May 2011 \\
\hline
\end{tabular}
\end{center}

\noindent
$^1$ W01: Weiss et al. (2001b); G01: Garc\'{\i}a-Burillo et al. (2001); F08: Fuente et al. (2008)
\label{tableObservations}
\end{table*}

The presence of a stellar bar in M~82 has been invoked as a mechanism 
to fuel the star formation activity in the inner $r\sim30\arcsec$ (500~pc) disk 
of the galaxy.  
The stellar bar, studied in detail by Greve et al.~(2002), could have 
formed during the encounter of
M~81 and M~82. The orbits sustaining the bar potential, denoted as $x_1$ 
orbits, are oriented parallel to the bar major axis, and extend out to 
$r\sim30\arcsec$ (500~pc) in the disk. The M~82's stellar bar seems to have 
room for $x_2$ orbits, which are oriented parallel to the bar minor axis, in 
the inner $r\sim 5\arcsec$($\sim$90~pc). Evidence supporting the existence of 
$x_2$ orbits in M~82 is derived from the position-velocity diagrams obtained 
in several tracers of the ISM, including CO, H{\small I} and [NeII] emission 
line data (Shen \& Lo~1995; Neininger et al.~1998; Wills et 
al~2000; Achtermann \& Lacy~1995).
Coupled stellar population synthesis and photoionization models reveal two major episodes of star formation over the past 
10~Myr (F\"orster-Schreiber et al. 2003). 
The first episode ($\sim$10~Myr) took place throughout the central regions of M~82 and was particularly intense
at the nucleus while the second episode  ($\sim$5~Myr) occurred predominantly in a circumnuclear ring at a radius of $\sim$90 pc 
and along the stellar bar. This inside-outside scenario can be understood as resulting from the gravitational interaction between
M~82 and its neighbour M~81, and subsequent bar-driven evolution. Each episode lasted only for a few million years suggesting
strong negative feedback effects from the starburst activity. 

X-rays and optical observations have shown the existence of a biconical outflow of hot gas 
coming out of the plane in the nucleus of M~82 (Bregman et al. 1995, Shopbell et al. 1998).
It is widely accepted that the driving mechanism of the outflow phenomenon in starbursts is 
linked to the creation of expanding shells of hot gas by supernovae. These hot bubbles blow 
out into the halo, entraining surrounding cold gas and dust at several hundreds of kilometers
per second.  
Interferometric observations showed that, contrary to most species, 
the SiO emission is not tracing the molecular gas in the galaxy disk. 
The detection of a $\sim$500 pc molecular gas chimney and 
a super-shell in SiO indicates the occurrence of large-scale shocks in the disk-halo
interface (Garc\'{\i}a-Burillo et al. 2001). Out-of-plane molecular emission has also
been detected in CO, HCN and HCO$^+$ (Weiss et al. 2001b, Salas et al. 2014). Recent observations of the HCN 1$\rightarrow$0 and
HCO$^+$ 1$\rightarrow$0 lines by Salas et al. (2014) showed that $>$2\% of the total dense molecular gas is in the outflow
formed by the gas expelled by the central starburst.

M~82 is one of the finest examples of how chemistry can help to the full understanding
of the interstellar medium (ISM) of an external galaxy. Our early HCO interferometric map using the PdBI showed that
the M~82 nucleus is a giant photon-dominated region (PDR) of $\sim$650~pc size.
Furthermore the comparison between the HCO and H$^{13}$CO$^+$ images suggested that
the PDR chemistry is propagating across the M~82 nucleus. Our subsequent 30m and PdBI chemical studies 
provided further support for the existence of giant PDRs in the galaxy disk (Garc\'{\i}a-Burillo et al. 2002, Fuente et al. 2005, 2006, 2008). 
In particular it is remarkable the detection of the reactive ion CO$^+$
using the 30m telescope (Fuente et al. 2006), the first time in an external galaxy, and the interferometric image of the
HOC$^+$ 1$\rightarrow$0 line (Fuente et al. 2008), the first one ever observed. 
Chemical modeling using the Meudon PDR code showed that most of the observations
(CO$^+$, HOC$^+$, HCO$^+$, CN, HCN, H$_3$O$^+$) are well explained assuming that about 87\%
of the mass of the molecular gas is forming small ($\sim$0.02~pc) clouds with a total 
thickness of $\sim$5 mag. In these small clouds all the molecular gas is exposed to an intense UV field and
the entire cloud is a PDR. Further star formation is not expected in this PDR component.
A small mass fraction ($\sim$13\%) must, however, be located in shielded regions 
to account for the measured [CN]/[HCN] abundance ratio.
We estimated that these clouds could have large column densities, with a total visual extinction of $>$50 mag.
The existence of this shielded component is also necessary to explain the detection of some species like NH$_3$ and CH$_3$OH 
(Weiss et al. 2001a, Mart{\'{\i}}n et al. 2006, Aladro et al. 2011). 

However, this chemical study was based on
single-dish 30m observations towards one position (position E in the nomenclature of Fuente et al. 2008) and cannot 
be extrapolated to the entire galaxy. In addition, all the observed molecules were PDR tracers which are
not adequate to trace the shielded gas. In this paper,
we present high spatial resolution ($\sim$100~pc) observations of a set of molecular lines
carried out with the Plateau de Bure Interferometer (PdBI). This set of molecules includes PDR tracers like CN
and c-C$_3$H$_2$ but also well-known tracers of cold and dense gas like N$_2$H$^+$, CS and C$^{18}$O. 
We combine the information from all these molecules to gain insights 
into the spatial distribution, physical conditions and chemistry of the 
molecular gas reservoir in the M~82 starburst.


\begin{figure} 
 \centering
 \includegraphics[width=0.45\textwidth] {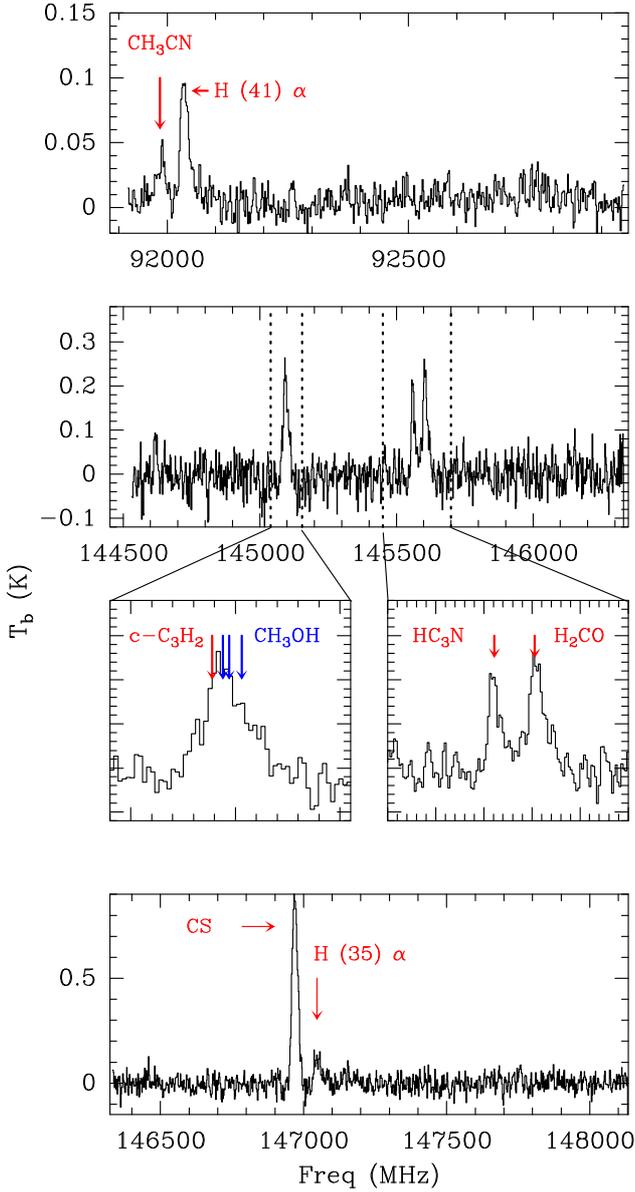}
 \caption{Interferometric spectra towards E1 [offset (+14\arcsec, +5\arcsec)] in temperature units (K). The velocity has been
set to v$_{lsr}$=330~km~s$^{-1}$, based on the C$^{18}$O data. Vertical lines indicate the following 
frequencies: 91987~(CH$_3$CN 5$_0$$\rightarrow$4$_{0}$), 
92034~(H(41)$\alpha$), 145561~(HC$_3$N 16$\rightarrow$15), 145603~(H$_2$CO 2$_{0,1}$$\rightarrow$1$_{0,1}$), 
145089~(c-C$_3$H$_2$ 3$_{1.2}$$\rightarrow$2$_{2,1}$), 145094~(E-CH$_3$OH 3$_{0,0}$$\rightarrow$2$_{0,0}$), 
145097~(E-CH$_3$OH 3$_{-1,0}$$\rightarrow$2$_{-1,0}$), 145103~(A-CH$_3$OH 3$_{0,0}$$\rightarrow$2$_{0,0}$), 
146969~(CS 3$\rightarrow$2), and 147047~(H(35)$\alpha$)~MHz.}
 \label{fig1}
\end{figure}

\section{Observations and data reduction}

The interferometric observations of the CN 1$\rightarrow$0 (113.491~GHz), 
N$_2$H$^+$ 1$\rightarrow$0 (93.173~GHz), CS 3$\rightarrow$2 (146.969~GHz), 
c-C$_3$H$_2$ 3$_{1,2}$$\rightarrow$2$_{2,1}$ (145.089~GHz), H$_2$CO 2$_{0,2}$$\rightarrow$1$_{0,1}$ (145.603~GHz) and 
HC$_3$N 16$\rightarrow$15 (145.601~GHz) lines were carried out with the IRAM Plateau de Bure Interferometer (PdBI)
as part of two different projects. The CN 1$\rightarrow$0 line was imaged in May and December, 2005,
with the antennas arranged in the C and D configurations providing almost a circular beam of 
2.46$\arcsec$$\times$2.27$\arcsec$~PA~170$^\circ$. The primary beam of the PdBI at this frequency is $\sim$44$''$,
enough to cover the whole galactic plane at 3mm and only one field was observed.
During the observations we adjusted the spectral correlator to give a contiguous bandwidth of 1~GHz with a frequency 
resolution of 2.5~MHz. The continuum maps were generated using the channels free of line emission. Then we subtracted 
the continuum emission to produce the spectroscopic maps. The maps  
have been corrected for the primary beam attenuation (primary beam=44$\arcsec$ at 113.5~GHz). 

In the second project we imaged the N$_2$H$^+$ 1$\rightarrow$0 (93.173~MHz), CS 3$\rightarrow$2 (146.969~MHz), 
c-C$_3$H$_2$ 3$_{1,2}$$\rightarrow$2$_{2,1}$ (145.089~MHz), H$_2$CO 2$_{0,2}$$\rightarrow$1$_{0,1}$ (145.603~MHz) and 
HC$_3$N 16$\rightarrow$15 (145.601~MHz) lines. These observations were carried out on December, 2010 and May, 2011,
with the antennas arranged in the C configuration.
During the 3mm observations, the narrow band correlators were placed to cover the N$_2$H$^+$ 1$\rightarrow$0 line with a 
frequency resolution of 20 kHz. 
We used the wide-band correlator WideX to cover a bandwidth of 3.6~GHz in dual polarization with a 
channel spacing of about 1.95~MHz. This allowed us to observe the H(41)$\alpha$ recombination line at 92.034~GHz 
and the CH$_3$CN 5$_k$$\rightarrow$4$_k$ k=0,1,2,3 at $\sim$91.987~GHz 
simultanously with the N$_2$H$^+$ 1$\rightarrow$0 line (see Fig, 1). Only one field was observed and
the maps have been corrected for primary beam attenuation (primary beam=54$\arcsec$ at 92.5~GHz). The synthesized beam was almost 
circular with HPBW$\sim$ 3.56$\arcsec$$\times$2.74$\arcsec$~PA~94$^\circ$. Because of the slightly better sensitivity,
we used the WideX cube for our analysis.

At 2mm, we observed two fields at (+7$\arcsec$, +2.5$\arcsec$) and (-7$\arcsec$, 
-2.5$\arcsec$). The CS 3$\rightarrow$2 line was covered with the narrow band correlators providing a 
frequency resolution of 160 kHz. The wide-band correlator WideX was used to cover a bandwidth of 3.6~GHz in dual polarization with a 
channel spacing of about 1.95~MHz. This allowed us to observe the H(35)$\alpha$ and the
c-C$_3$H$_2$ 3$_{1,2}$$\rightarrow$2$_{2,1}$, CH$_3$OH 3$\rightarrow$2, H$_2$CO 2$_{0,2}$$\rightarrow$1$_{0,1}$ and 
HC$_3$N 16$\rightarrow$15 lines in the same setting. In Fig.~\ref{fig1} we present the observed spectrum towards
the position (+14$\arcsec$,+5$\arcsec$). Unfortunately, the c-C$_3$H$_2$ line
is blended with several CH$_3$OH 3$\rightarrow$2 lines
and cannot be separated (see Fig.~\ref{fig1}). The H$_2$CO 
2$_{0,2}$$\rightarrow$1$_{0,1}$ (145.603~MHz) and HC$_3$N 16$\rightarrow$15 (145.601~MHz) lines are 
partially blended and cannot be imaged separately, but can be separated in the outer part of the galaxy where the 
linewidths are narrower. We  generated continuum maps using the channels free of line emission to subtract 
the continuum from the spectral images. The HPBW of the synthesized beam is $\sim$2.3$\arcsec$$\times$1.88$\arcsec$~PA~75$^\circ$.

Data analysis and image processing have been done using the GILDAS package software (http://www.iram.fr/IRAMFR/GILDAS/)
and JPARSEC (http://conga.oan.es/~alonso/doku.php?id=jparsec).

\begin{figure*} 
 \centering
 \includegraphics[width=0.8\textwidth] {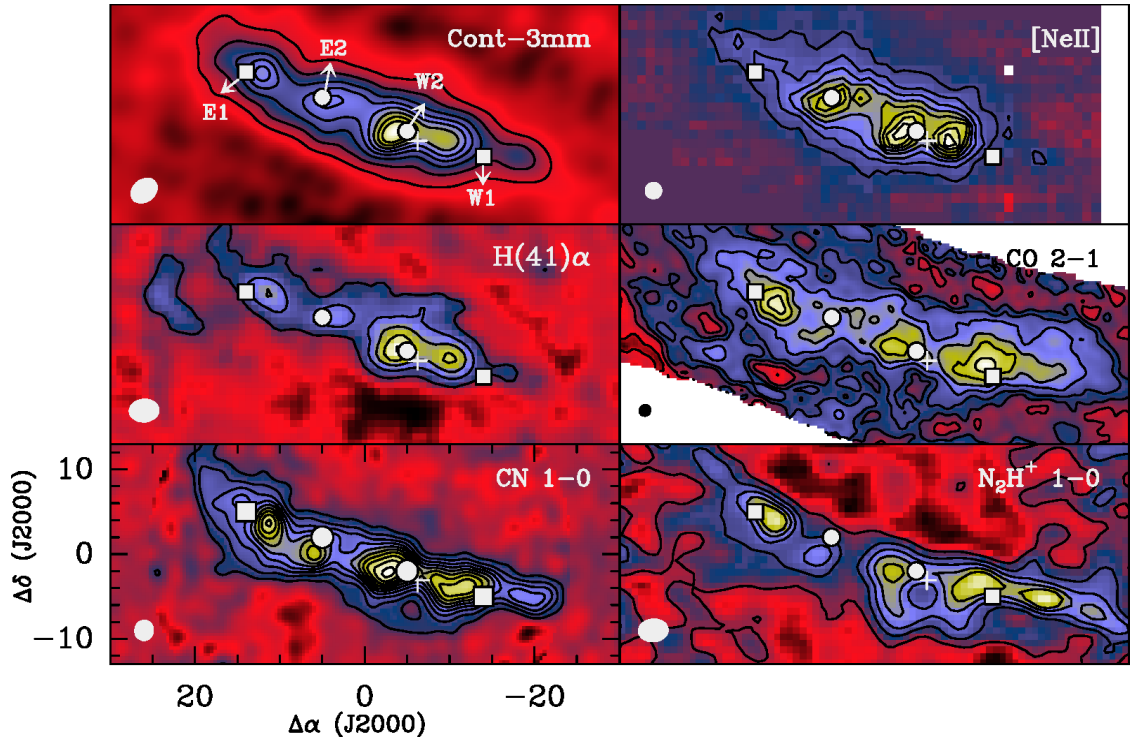}
 \caption{Map of the continuum emission at 3mm (Fuente et al. 2008) and line integrated intensity maps of the
12.8~$\mu$m  line of [NeII] (Achtermann \& Lacy 1995), H(41)$\alpha$, CO 2$\rightarrow$1, 
CN 1$\rightarrow$0 and N$_2$H$^+$ 1$\rightarrow$0 lines. 
The beam is drawn in the left-down corner of the panels and positions E1, E2, W2 and W1 are indicated by filled
polygons. The position of SN 41.9+58 is marked with a white cross. 
First contour is at 3$\times$$\sigma$ level. Contours levels are: 
1.7~mJy/beam, 5~mJy/beam to 35~mJy/beam in steps of 5~mJy/beam (cont-3mm); 
0.13~Jy/beam to 1.04~Jy/beam in steps of 0.13~Jy/beam ([NeII]);
4 to 28 in steps of 4~K~$\times$~km~s$^{-1}$ (H(41) $\alpha$);
200, 400, 600, 800, 1200, 1600, 2000, 2400~K $\times$~km~s$^{-1}$ (CO);
15.8 to 252.8 in steps of 15.8~K~$\times$~km~s$^{-1}$ (CN); 
2.5, 5.0, 7.5, 10.0, 12.5~K~$\times$~km~s$^{-1}$ (N$_2$H$^+$). Colour scale is adjusted to the minimum 
and maximumm values in each panel.}
 \label{fig2}
\end{figure*}

\begin{figure*} 
 \centering
 \includegraphics[width=0.8\textwidth] {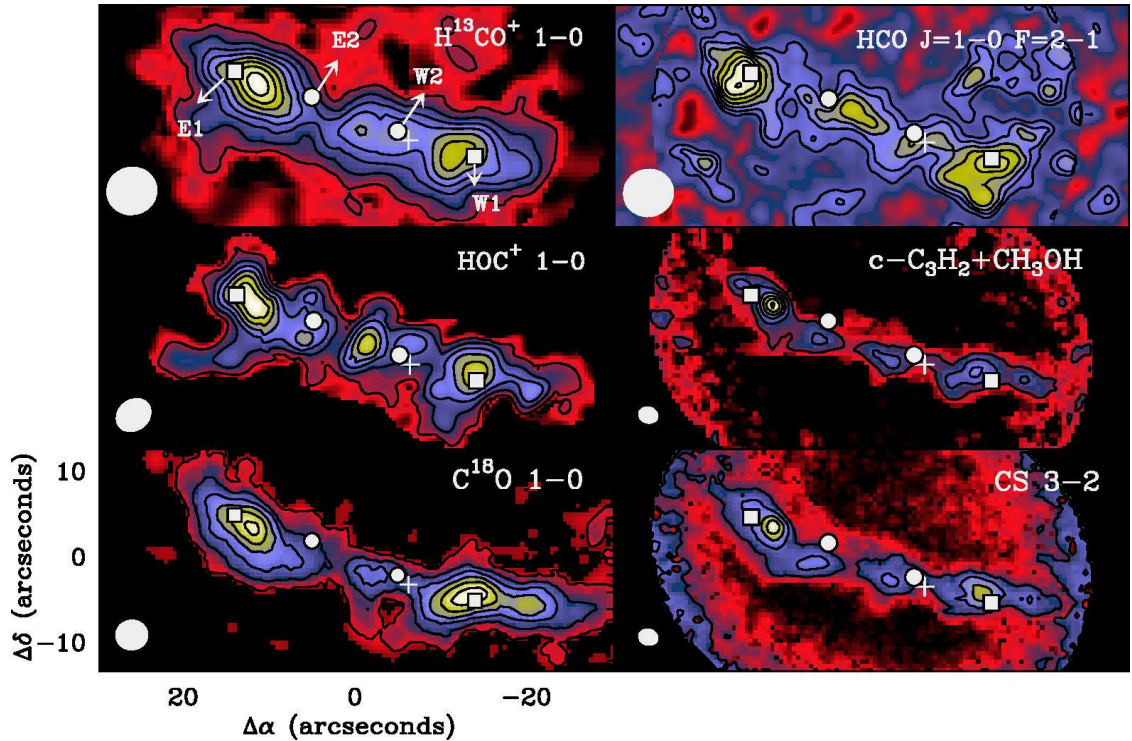}
 \caption{The same as Fig. 2 for the H$^{13}$CO$^{+}$ 1$\rightarrow$0, HCO 1$\rightarrow$0 F=2$\rightarrow$1, 
 HOC$^+$ 1$\rightarrow$0, C$_3$H$_2$ 3$_{1,2}$$\rightarrow$2$_{2,1}$, C$^{18}$O 1$\rightarrow$0 and  CS 3$\rightarrow$2
 lines. Contour levels are:  
0.6 to 4.6 in steps of 0.6~K~$\times$~km~s$^{-1}$ (H$^{13}$CO$^{+}$); 
0.7 to 1.9 in steps of 0.2~K~$\times$~km~s$^{-1}$ (HCO);
0.28, 0.70, 1.12,1.54, 1.96, 2.38~K~$\times$~km~s$^{-1}$ (HOC$^+$); 
6.4 to 44.8 by steps of 6.4~K~$\times$~km~s$^{-1}$ (C$_3$H$_2$); 
2, 8, 14, 20, 26, 32~K~$\times$~km~s$^{-1}$ (C$^{18}$O);
30 to 150 in steps of 30~K~$\times$~km~s$^{-1}$ (CS). }
 \label{fig3}
\end{figure*}

All the maps are centered at, RA = 09:55:51.9, Dec = +69:40:47.10 (J2000), that is 
the 2.2~$\mu$m peak as determined by Joy, Lester \& Harvey (1987). This is also the central position used in previous 
interferometric studies (Weiss et al. 2001b, Garc\'{\i}a-Burillo et al. 2001, 2002, Fuente et al. 2008).
Lester et al. (1990) established the 2~$\mu$m peak at RA = 09:55:52.4, Dec = +69:40:46.00 (J2000),
i.e. +2.7$\arcsec$ E 1.3$\arcsec$ S from our center position. The M~82 galactic plane is viewed
almost edge-on (i$\sim$80$^\circ$, Greve 2011). In order to guide the discussion, we have defined 
four positions E1, E2, W2, and W1 [(+14$\arcsec$,+5$\arcsec$), (+5$\arcsec$,+2$\arcsec$),
($-$5$\arcsec$,$-$2$\arcsec$) and ($-$14$\arcsec$,$-$5$\arcsec$), respectively]
that are aligned with the center at an angle of 68$^\circ$. 
Within the errors, this is the position angle of the galactic plane.
The intense continuum source SN~41.9+58 is located ($-$1.25$\arcsec$,$-$1.1$\arcsec$) from our position W2
(Kronberg et al. 1981, Weiss et al. 1999). Position E1 coincides with position E in Fuente et al. (2008).
The stellar bar is inclined $\sim$4$^\circ$ relative to the galactic plane, with the western side above the adopted major
axis of the disk.
This explains the tilted appearance in the molecular emission (see Figs.~\ref{fig2} and \ref{fig3}).

In addition to these new observations, we use previous interferometric data published 
by Garc\'{\i}a-Burillo et al. (2001), Fuente et al. (2008) and
Weiss et al. (2001b). In Table 1, we give a brief description of the observations and their more relevant observational parameters.

\begin{figure} 
 \centering
 \includegraphics[width=0.4\textwidth] {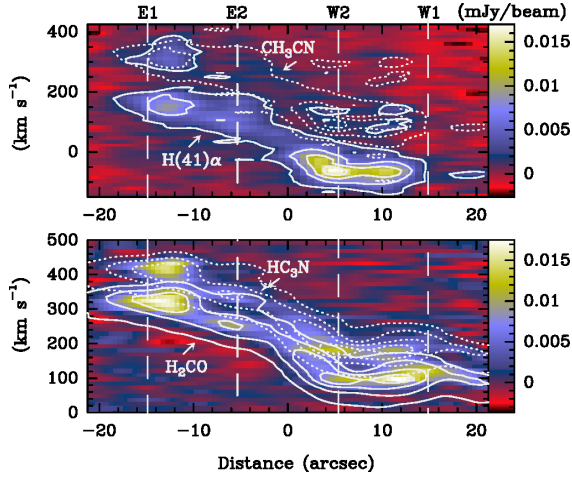}
 \caption{{\it Top:} In colour scale, the position-velocity (p-v) diagram along the galactic plane 
 (crossing E1, E2, W2 and W1) of H(41)$\alpha$. Solid contours are: 
 2.1 mJy/beam to 16.6 mJy/beam by 4.2 mJy/beam. In dashed lines, the same contours but centered 
 at the frequency of CH$_3$CN 5$_0$$\rightarrow$4$_0$ line (91.987 GHz).
 {\it Bottom:} In colour scale, the P-V diagram of the H$_2$CO 2$_{0,1}$$\rightarrow$1$_{0,1}$ line. 
 For comparison, we have drawn in solid contours the levels of the CN 1$\rightarrow$0 line: 
10.0 mJy/beam to 86.6 mJy/beam by 20.0~mJy/beam. In dashed lines, the same contours at the frequency of
the HC$_3$N 16$\rightarrow$15 line. Note that the H$_2$CO 2$_{0,1}$$\rightarrow$1$_{0,1}$) and 
HC$_3$N 16$\rightarrow$15 lines are heavily blended in the inner region.}
 \label{fig4}
\end{figure}

\section{Results}
Fig.~\ref{fig1} shows the wide band spectra towards E1.
The velocity integrated intensity maps for all the lines detected are shown in Figs.~\ref{fig2} and \ref{fig3}. 
The maps of the H$_2$CO  2$_{0,2}$$\rightarrow$1$_{0,1}$ (145.603~MHz) and 
HC$_3$N 16$\rightarrow$15 (145.601~MHz) are not shown
because the lines cannot be deblended in the inner region. In Fig.~\ref{fig4}, we show the
position-velocity (p-v) diagrams along the galactic plane (straight line across E1, E2, W2 and W1)
of the H$_2$CO  2$_{0,2}$$\rightarrow$1$_{0,1}$, HC$_3$N 16$\rightarrow$15, CH$_3$CN 5$_k$$\rightarrow$4$_k$ 
and H(41)$\alpha$ lines. The CH$_3$CN 5$_k$$\rightarrow$4$_k$ is only detected towards the outer part of 
the galaxy and, at 3$\sigma$ level, around W2. 
We subtracted the CH$_3$CN 5$_k$$\rightarrow$4$_k$ line from the data cube to create the H(41)$\alpha$ 
map shown in Fig.~\ref{fig2}.
In Fig.~\ref{fig3}, we show the c-C$_3$H$_2$ map although we are
aware that the line is contaminated with CH$_3$OH emission. We will discuss this problem in Sect.~4.

\begin{figure} 
 \centering
 \includegraphics[width=0.4\textwidth] {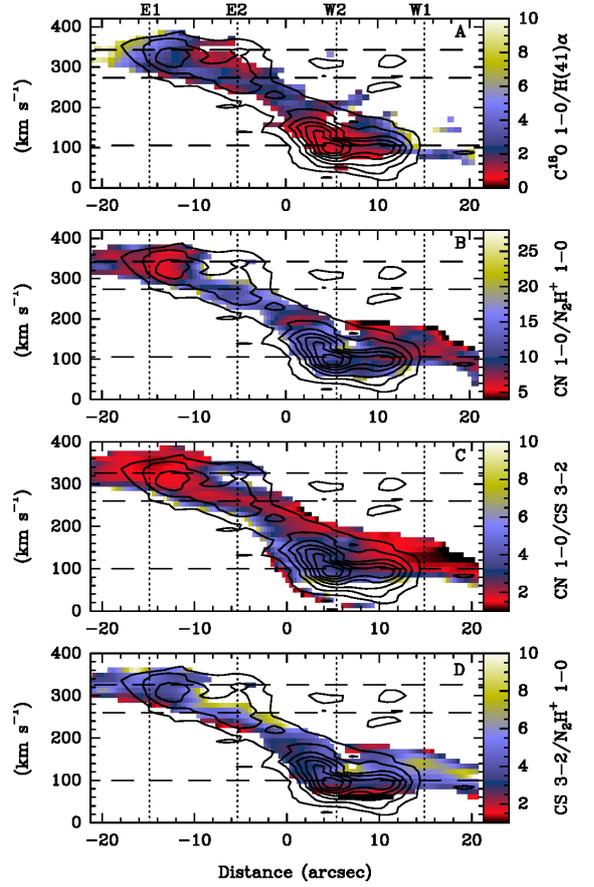}
 \caption{P-v diagrams along a cut crossing E1, E2, W2 and W1. Contours correspond to the line intensity of H(41)$\alpha$
(2.1 to 16.6 in steps of 2.1 mJy/beam). Colour scale shows the following line intensity ratios (in T$_b$ units): 
{\bf A)} C$^{18}$O 1$\rightarrow$0/H(41)$\alpha$;
{\bf B)} CN 1$\rightarrow$0/N$_2$H$^+$ 1$\rightarrow$0; {\bf C)} CN 1$\rightarrow$0/[CS] 3$\rightarrow$2;
{\bf D)} CS 3$\rightarrow$2/N$_2$H$^+$ 1$\rightarrow$0. Horizontal lines correspond to velocities: 100, 270 and 326 km~s$^{-1}$, to
make easier the comparison with Fig.~7. Line ratios have been calculated with the original images degraded to common angular 
resolution of 3.8$\arcsec$ and a velocity resolution of 12~km~s$^{-1}$. The C$^{18}$O 1$\rightarrow$0/H(41)$\alpha$ line ratio can
be considered as an observational measurement of the gas illumination by UV photons.
Note that the CN/N$_2$H$^+$ and CN/CS line ratios increase in the highest UV irradiated regions, i.e., minimum
values of the C$^{18}$O 1$\rightarrow$0/H(41)$\alpha$ line ratio.}
 \label{fig5}
\end{figure}

\begin{figure} 
 \centering
 \includegraphics[width=0.4\textwidth] {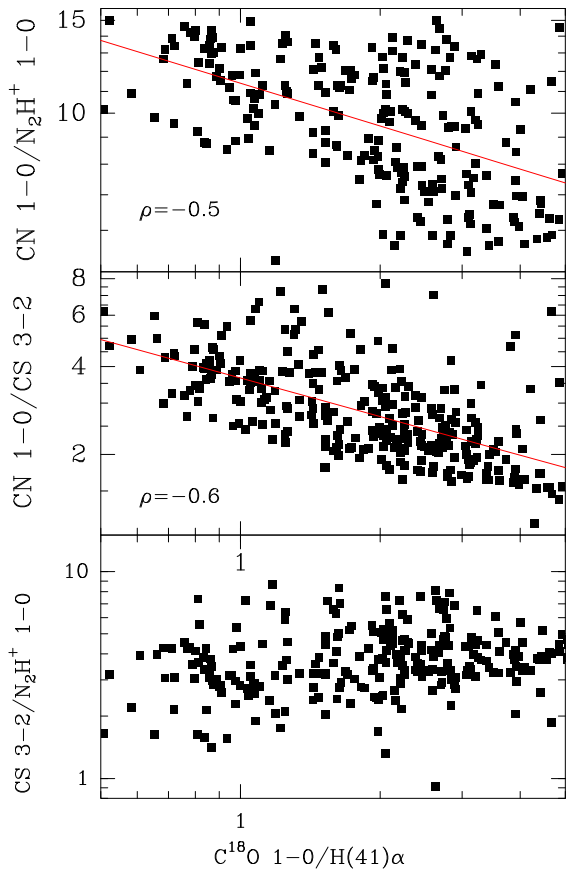}
 \caption{Correlations between the  C$^{18}$O 1$\rightarrow$0/H(41)$\alpha$, CN 1$\rightarrow$0/N$_2$H$^+$ 1$\rightarrow$0,
and  CS 3$\rightarrow$2/N$_2$H$^+$ 1$\rightarrow$0 brighness temperature line ratios along the galactic plane shown in Fig.~5. 
The plots are in logarithmic scale. In order to get a more reliable result avoiding the line borders, we only considered the points 
within the following range of values: 0.5 and 5 for 
C$^{18}$O 1$\rightarrow$0/H(41)$\alpha$ line; 5 and 15 for CN 1$\rightarrow$0/N$_2$H$^+$ 1$\rightarrow$0; 1 and 6 
for CN 1$\rightarrow$0/CS 3$\rightarrow$2 between 1 and 6. The results of the least square fitting 
and the values of the Pearson correlation coefficients are shown in the top and medium panels.}
 \label{fig6}
\end{figure}

\subsection{Galactic plane}

The emissions of all the species studied in this paper are concentrated in the galactic plane.
However, there are some differences in their spatial distributions that we below discuss.

In Fig.~\ref{fig2} we show the integrated intensity maps of the lines that present intense emission
towards the inner positions, E2 and W2. 
E2 and W2 are related with the ring of ionized gas that is associated with the most 
recent starburst ($\sim$5~Myr) (F\"orster-Schreiber et al. 2011),
This ring is asymmetric with the western part more intense than the eastern one.
The continuum emission at 3mm and the ionized line [NeII] are maxima towards 
W2 (see Fuente et al. 2008 and Fig.~\ref{fig2}). As expected, the emission of H(41)$\alpha$ follows
the same spatial distribution. The emissions of the molecular lines are, however, more intense
in the outer part of the galaxy close to positions E1 and W1, with the only exception of
CN that presents its emission peak towards the inner ring
following the spatial distribution of the ionized lines. 
CN is known to present higher abundances in regions with enhanced UV fields
(Fuente et al. 1993, 1995; Bachiller et al. 1997;
Boger \& Sternberg 2005). Its peculiar spatial distribution
suggests that the UV radiation from the recently formed stars has a strong impact on the 
molecular gas chemistry in the inner ring.

The emission of most molecular lines, H$^{13}$CO$^+$ 1$\rightarrow$0, HOC$^+$ 1$\rightarrow$0, 
C$^{18}$O 1$\rightarrow$0, HCO 1$\rightarrow$0, C$_3$H$_2$ 3$_{1,2}$$\rightarrow$2$_{2,1}$, and CS 3$\rightarrow$2
is, however, more intense towards the outer positions, E1 and W1 (see Fig.~\ref{fig3}). Although all these molecules are 
brighter in the outer part of the galaxy, the exact positions of their emission
peaks differ, especially in the western part. The C$_3$H$_2$ 3$_{1,2}$$\rightarrow$2$_{2,1}$, 
CS 3$\rightarrow$2, H$^{13}$CO$^+$ 1$\rightarrow$0, and HOC$^+$ 1$\rightarrow$0 peaks are located  2$''$--4$''$
closer to the dynamical center of the galaxy than HCO 1$\rightarrow$0. The intense peak
of the c-C$_3$H$_2$ line towards E1 is very likely due to the CH$_3$OH contamination (see Sect.~4).

In Fig.~\ref{fig5} we show the p-v diagrams of
the CN 1$\rightarrow$0/N$_2$H$^+$ 1$\rightarrow$0, CN 1$\rightarrow$0/CS 3$\rightarrow$2 and the
CS 3$\rightarrow$2/N$_2$H$^+$ 1$\rightarrow$0 line ratios along the plane defined
by positions E1, E2, W1 and W2.
To perform these diagrams, we have degraded the spatial and spectral resolutions of the three 
line intensity cubes to common values of 3.8$\arcsec$ and 12 km s$^{-1}$, respectively.
We adopted 3$\times$$\sigma$ as the threshold for line detection. 
There is a clear gradient in the CN 1$\rightarrow$0/N$_2$H$^+$ 1$\rightarrow$0 line ratio along the
galaxy plane, being a factor of $\sim$3 higher in the inner region than in E1. 
For comparison, we show the p-v diagram of the C$^{18}$O 1$\rightarrow$0 /H(41)$\alpha$ line 
ratio that can be considered as an observational tracer of the gas illumination.
There are two regions where the CN 1$\rightarrow$0/N$_2$H$^+$ 1$\rightarrow$0 takes the
peak value of $\sim$15: (i) W2, it is the peak 
in the H(41)$\alpha$ line emission and the region with the highest concentration
of HII regions; and (ii) E2, the emission of the H(41)$\alpha$ line is lower than in W2
but the molecular gas column density is also lower, especially in 
the $\sim$326~km~s$^{-1}$ component that is not detected in N$_2$H$^+$ (see Sect. 3.2); 
as a consequence, the molecular gas is very likely immersed in a high UV field (see
Fig.~\ref{fig5}, A).
The CN 1$\rightarrow$0/CS 3$\rightarrow$2 line ratio has a ratio $\sim$1$-$2 along the cut and
increases to $\sim$5 only in the two enhanced UV field regions described 
above (see Fig.~\ref{fig5}, C). 
The variations of these two molecular line ratios
are not hazardous but seems related with the distribution of the ionized gas 
suggesting that UV photons are the driving
agent of the molecular chemistry. 
In Fig.~\ref{fig6} we show the log-log correlation diagrams between the CN 1$\rightarrow$0/N$_2$H$^+$ 1$\rightarrow$0
and  CN 1$\rightarrow$0/CS 3$\rightarrow$2 line ratios, and C$^{18}$O 1$\rightarrow$0 /H(41)$\alpha$. As expected
there is a good (anti-)correlation between these ratios at low extinctions (low values of C$^{18}$O 1$ç\rightarrow$0 /H(41)$\alpha$).
At high extinctions, the dispersion is larger since we can have PDRs (diffuse clouds, the external layers of giant molecular
clouds, dense gas around Herbig Ae/Be stars) that do not emit in H(41)$\alpha$ 
but also present enhanced CN abundances (see e.g. Fuente et al. 1993, 1995, Liszt et al. 2001).
On the contrary, although with a large dispersion, the
CS 3$\rightarrow$2/N$_2$H$^+$ 1$\rightarrow$0 line ratio is $\sim$5 towards all positions. 

\begin{figure*} 
 \centering
 \includegraphics[width=0.9\textwidth] {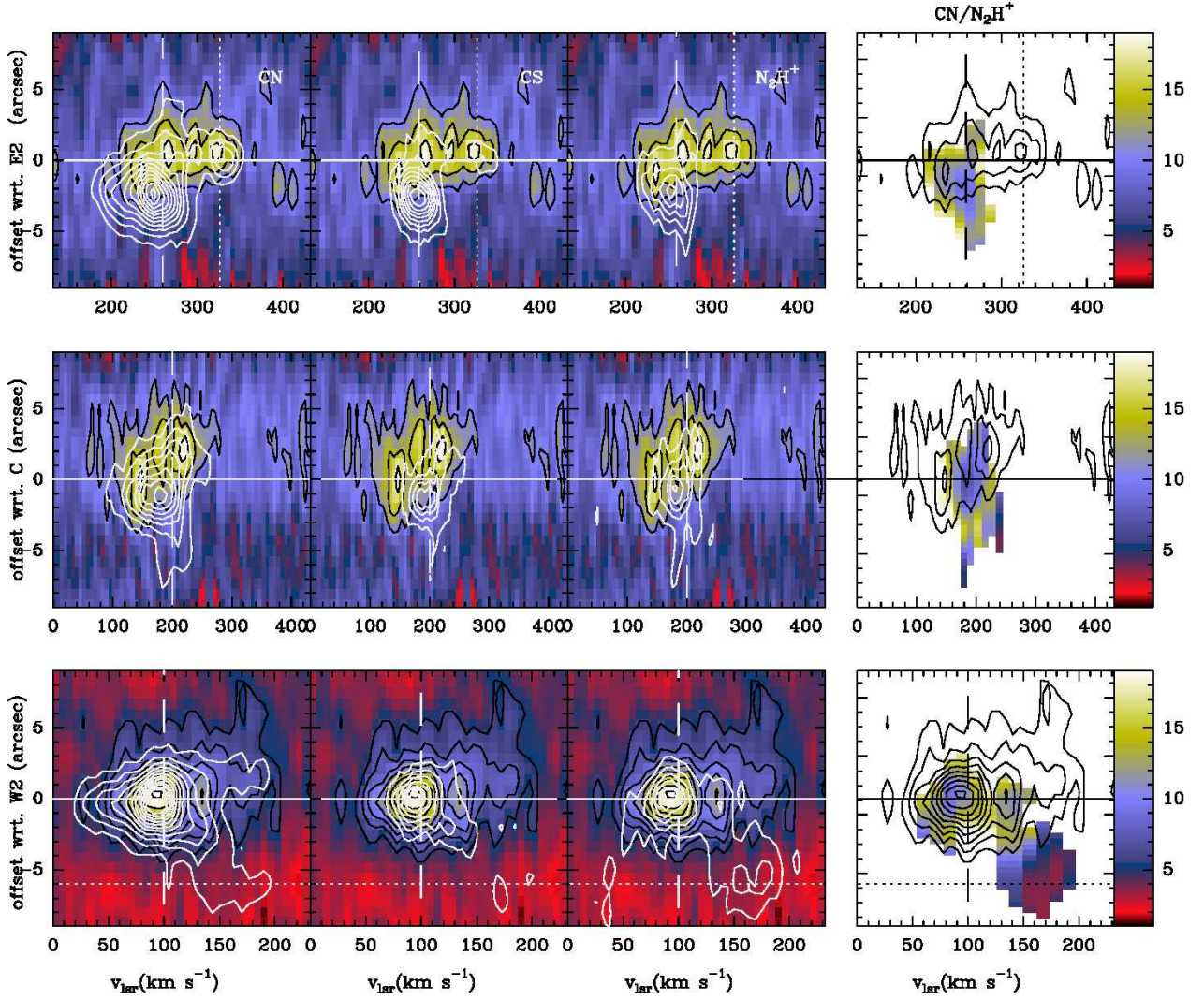}
 \caption{P-v diagrams along cuts perpendicular to the galactic plane at E2 (top), the (0,0) position (middle) and W2 (bottom). 
Colour scale and black contours correspond to the H(41)$\alpha$ emission. White contours are intensities of the CN 1$\rightarrow$0, CS 3$\rightarrow$2 and 
N$_2$H$^+$ 1$\rightarrow$0 lines. First contours (3$\times$$\sigma$) and steps are: 10 mJy/beam for CN and CS; 2 mJy/beam for N$_2$H$^+$ and H(41)$\alpha$. 
The CN 1$\rightarrow$0/N$_2$H$^+$1$\rightarrow$0 line intensity (in T$_b$ units) ratio along these cuts is shown in the right column. The images 
have been degraded to a angular resolution of 3.8$\arcsec$ and a velocity resolution
of 12~km~s$^{-1}$ to perform this line ratio. Vertical lines indicate of the position of the velocity components discussed in the text. }
 \label{fig7}
\end{figure*}

\subsection{Vertical distribution of the molecular emission}

M~82 is associated with a kpc-scale outflow with the gas coming towards us in the northern part. Optical and near infrared 
lines look asymmetric with the most intense half towards the north where the
radio continuum chimneys are located (Wills et al. 1999). This asymmetry is also detected in the emission of H(41)$\alpha$ 
which proves that it is not a consequence of the dust extinction but to the spatial distribution of the massive star forming regions 
located in the northern half of the disk (see Fig.~\ref{fig7}). 
On the contrary, the molecular emission extends to $\sim$7.5$\arcsec$ south 
from the galactic plane following the molecular supershell around the supernova remnant SN~41.9+58. 
Regarding the molecular emission, there are important differences between the 
vertical spatial distribution of the different species. 

One interesting case is the comparison of the HCO$^+$ and HOC$^+$ emissions. Fuente et al. (2008) found a clear north-south gradient in 
the [HCO$^+$]/[HOC$^+$] ratio being this ratio maximum towards the southern part of the supershell associated to SN~41.9+58. 
The maximum in the HOC$^+$ emission is 
shifted $\sim$2.5$\arcsec$ north relative to H$^{13}$CO$^+$, i.e. closer to the ionized layer traced by H(41)$\alpha$. Fuente et al. (2008) interpreted 
this gradient as a change in the global ionization degree of the molecular clouds. 

We used the CN, N$_2$H$^+$, CS and H(41)$\alpha$ images to make a detailed study of the kinematics and 
spatial distribution of the molecular gas in the direction perpendicular to the galaxy plane. In Fig.~\ref{fig7} we show the p-v
diagrams in the vertical direction across E2, the dynamic center of the galaxy, and W2. In these diagrams, the velocity axis has been
re-binned to channels of 12~km~s$^{-1}$. 

Towards E2, several velocity components 
are detected in the H(41)$\alpha$ and molecular lines. All these components are within the range of velocities of the $x2$ bar 
orbits, $\pm$120~km~s$^{-1}$ (Greve 2011).
The component at $\sim$260~km~s$^{-1}$ is the most intense in molecular emission and is detected in CN, N$_2$H$^+$ and CS.
However, the component at $\sim$326~km~s$^{-1}$ is well detected in the CN 1$\rightarrow$0 line, only tentatively detected in 
CS 3$\rightarrow$2 (3$\times$$\sigma$ level) and remains 
undetected in the N$_2$H$^+$ 1$\rightarrow$0 line. This component is associated with
a secondary peak in the H(41)$\alpha$ emission. In the right panel, we show the 
CN 1$\rightarrow$0 /N$_2$H$^+$1$\rightarrow$0  line intensity 
ratio. It varies between $\sim$10 and $\sim$15 along this cut with the peaks following the peaks of the H(41)$\alpha$ emission.

Several velocity molecular components are also detected towards W2. In this case, we can also observe a 
layered structure in the vertical direction with the H(41)$\alpha$ and CN emission being extended towards north 
while N$_2$H$^+$ emission is more extended towards the south. Again, there are important chemical differences between the 
$\sim$100 km s$^{-1}$ (north) and $\sim$160 km s$^{-1}$ (south) components. The north component is more intense in CN and CS, while 
the south component is brighter in the N$_2$H$^+$ emission. The CN 1$\rightarrow$0 /N$_2$H$^+$1$\rightarrow$0  
line intensity ratio shows an excellent correlation with the H(41)$\alpha$ emission, with a contrast of a factor of $\sim$5 between
the north and south components. The lowest values, $\sim$2$-$3, are found towards the south, and are even lower than those observed 
in the outer part (E1,W1) of the galaxy (see Fig.~\ref{fig5}).

For comparison, we also show the p-v diagram across the dynamic center
of the galaxy. 
The H(41)$\alpha$ emission is detected shifted towards the north compared with the molecular emission.
Similarly to the vertical cut across E2, the CN 1$\rightarrow$0 /N$_2$H$^+$1$\rightarrow$0  line intensity ratio vary between 
$\sim$10 and $\sim$15 with the peaks towards the peaks of the H(41)$\alpha$ emission, suggesting that it is a general trend all over
the galactic nucleus. 
 
\begin{figure*} 
 \centering
 \includegraphics[width=0.9\textwidth] {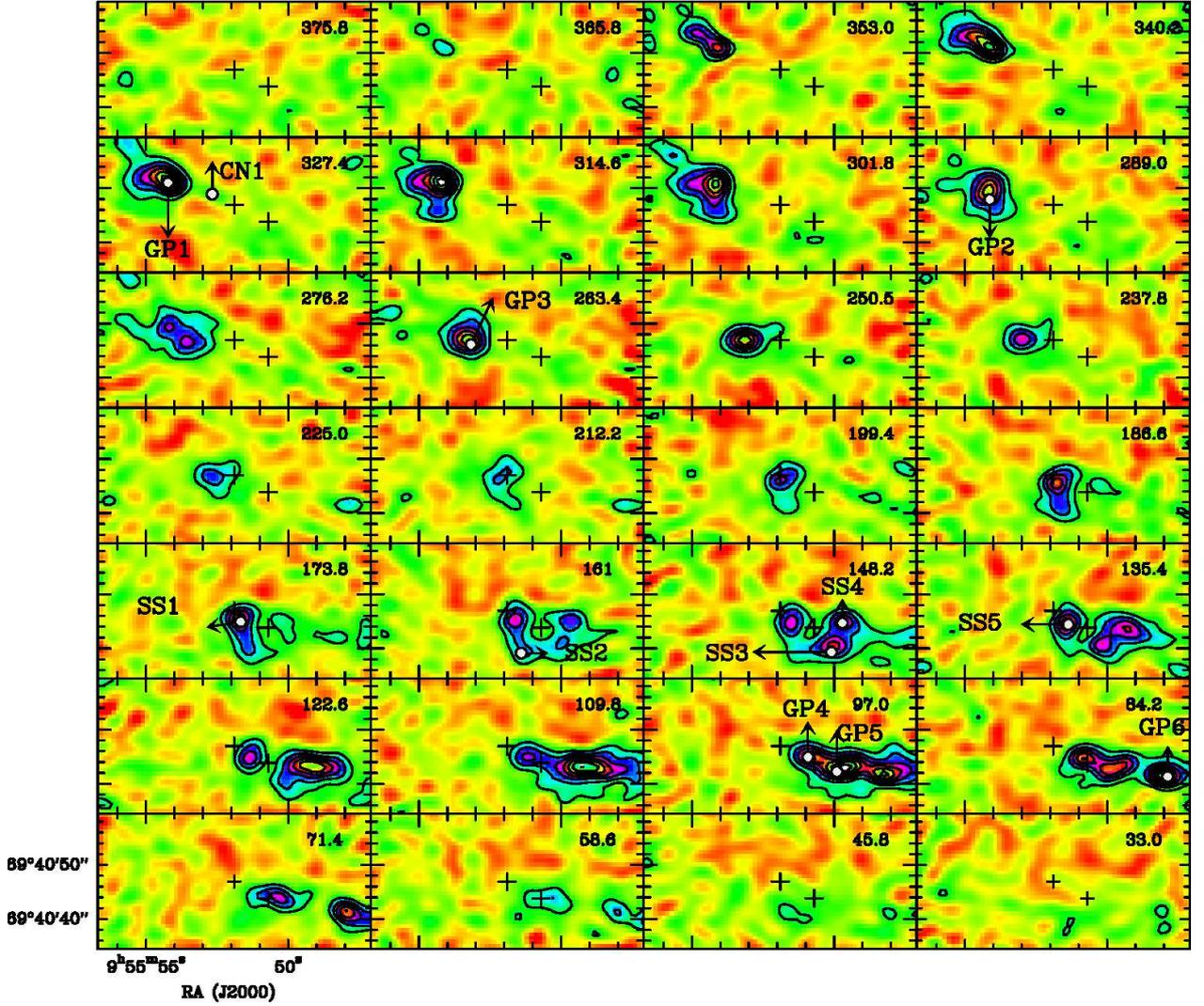}
 \caption{Spectral maps of the N$_2$H$^+$ 1$\rightarrow$0 line convolved to an angular resolution of 3.8$\arcsec$. 
 To increase the S/N ratio, the velocity resolution has been degraded to 12.8~km~s$^{-1}$. 
 The number in the right-upper corner indicates the central channel velocity. First contour and step are  
0.028~K ($\sim$3$\times$$\sigma$). Crosses indicate the dynamical center of the galaxy and 
the position of SN~41.9+58.
 }
 \label{fig8}
\end{figure*}

\section{Column densities towards selected positions}

\subsection{CN, N$_2$H$^+$, c-C$_3$H$_2$ and CS}
The high angular and spectral resolutions provided by our new data: CN 1$\rightarrow$0, N$_2$H$^+$ 1$\rightarrow$0, H(41)$\alpha$, CS 3$\rightarrow$2 and
C$_3$H$_2$ 3$\rightarrow$2, allow us to carry out a detailed chemical study of the interstellar medium in M~82. The spectral cubes of 
all these lines have been degraded to a common angular resolution of 3.8$\arcsec$ ($\approx$70~pc) and a channel 
width of 12.8~km~s$^{-1}$, that are typical values for the
size and linewidth of a giant molecular cloud (see e.g. Murray 2011). From these data, we have selected 12 molecular 
position-velocity peaks (see Fig.~\ref{fig8}) to carry out our 
chemical modeling. The selection criteria are: the knots must be (i) intense and (ii) well identified in velocity and position. 
In addition, the whole set must represent the variety of physical and chemical conditions in the galaxy. 
The intense and compact emission towards 
these positions minimizes the uncertainties due to spatial filtering effects.
The selected knots are listed Table 2 and indicated in Fig.~\ref{fig8} and Figs.~\ref{figa1} to \ref{figa3}.

GP1 to GP6 are compact molecular peaks along the galactic plane that follow the velocity pattern of the
$x1$ and $x2$ bar orbits. These knots have been detected in all molecular lines (see Fig.~\ref{fig8} and 
Figs.~\ref{figa1}-\ref{figa3}). 
We have added 5 positions, SS1 to SS5, placed in the supershell associated to SN~41.9+58. 
Note that SS1 and SS2 are very close to GP4 in the plane of sky and are only distinguishable 
because of their different velocities.
Examining the CN spectral maps, we identified one well defined CN emission peak with weak counterpart in the other 
species. We will refer to this position as CN1 and corresponds to the $\sim$326~km~s$^{-1}$ component
towards E2 (see Figs.~\ref{fig7} and ~\ref{figa1}). 
Fig.~\ref{figa3} shows the H(41)$\alpha$ PdBI spectral maps. This line probes the
spatial distribution of the ionized gas that is closely related to the local UV field. GP4 is the most intense position in 
this radio recombination line. Secondary peaks are spatially coincident with GP1, CN1 and GP5 showing the presence of
ionized gas at these positions. The absence of H(41)$\alpha$ emission towards SS2 and SS3 proves 
the existence of a lower UV field in the sourthern part of the supershell. Therefore, our selected positions are 
representative of the different physical conditions, in particular the mean UV field, within this galaxy.

We have estimated the CN, N$_2$H$^+$, CS and C$_3$H$_2$ molecular column densities using the LVG
code MADEX (Cernicharo 2012). Our observations do not allow a multitransitional study. 
Instead, we need to assume a uniform layer with constant physical conditions. On basis of CO and its isotopologues, 
Weiss et al. (2001b) derived densities of n(H$_2$)$\sim$5$\times$10$^4$~cm$^{-3}$ and gas kinetic 
temperatures ranging from $\approx$50~K towards the E1 and W1 positions to $\approx$150~K towards E2 and W2. 
Later, Fuente et al. (2005) determined densities of  
1$\times$10$^5$~cm$^{-3}$ from single-dish data of the CN 1$\rightarrow$0 and 2$\rightarrow$1 lines.
Higher densities, $\sim$5$\times$10$^5$~cm$^{-3}$, were estimated by Fuente et al. (2008) using the 
high excitation  J=3$\rightarrow$2 and 4$\rightarrow$3 lines of HCO$^+$.  
Bayet et al. (2009) proposed two components to fit the CS lines with densities of 
$\sim$10$^5$~cm$^{-3}$ and 6$\times$10$^5$~cm$^{-3}$ for the diffuse and dense components, respectively.
These estimates proved that, as expected, different density components coexist within our beam.
Since we are dealing with molecules with large dipole moments, we considered that the emission 
is dominated by dense gas with n(H$_2$)$>$1$\times$10$^5$~cm$^{-3}$. 
To estimate the molecular column densities and the uncertainties in our
calculations we run a grid of LVG models with n(H$_2$) varying from 1$\times$10$^5$-5$\times$10$^5$~cm$^{-3}$
and T$_k$=50, 150~K and adopted the average value. The hyperfine structures of CN and N$_2$H$^+$ have not been 
considered in our calculations because of the large linewidths. The estimated column densities are shown in Table~2. 
The errors correspond to the minimum and maximum values obtained in our grid, i.e., these errors do not account for 
the observational errors, which are lower, but for the uncertainty in the physical conditions. In general,
the uncertainties in the obtained column densities are within a factor of 2.

In our calculations we assumed that all the emission at 145.089 GHz comes from the c-C$_3$H$_2$ carrier
that is not true for some positions. Methanol was firstly detected by Mart\'{\i}n et al. (2006) 
in M~82. Based on the kinematical and spatial distribution, they concluded that the emission is mainly coming from
the intense east and west knots at the outer part of the galaxy. Later, Aladro et al. (2011) performed a single-dish
multitransitional study of CH$_3$OH anc c-C$_3$H$_2$ towards a position close to E1 and derived their column densities
and rotational temperatures. We assumed these values to predict the expected CH$_3$OH and c-C$_3$H$_2$ line intensities and 
obtained that $\sim$50\% of the emission must belong to each carrier towards E1. According to these results the 
c-C$_3$H$_2$ column density is overestimated by a factor of $\sim$2 in E1. Taking into account the similarties
between the molecular chemistry in E1 and W1, we consider that, very likely, we have a similar contamination, i.e. $\sim$50\%,
towards W1.

In Table 2, we also show the [CN]/[N$_2$H$^+$], [CS]/[N$_2$H$^+$] and [CN]/[CS] column density ratios towards the selected molecular knots.
These values have been estimated using the same procedure as for the column densities. We run a grid of LVG models and adopted the average 
value of the fitted column density ratios. 
When molecules with similar excitation conditions are selected, this method minimizes the uncertainty due to the assumed physical 
conditions, being the uncertainties in the column density ratios lower than those of the individual column 
densities (see Table~2).

\begin{figure*} 
 \centering
 \includegraphics[width=0.9\textwidth] {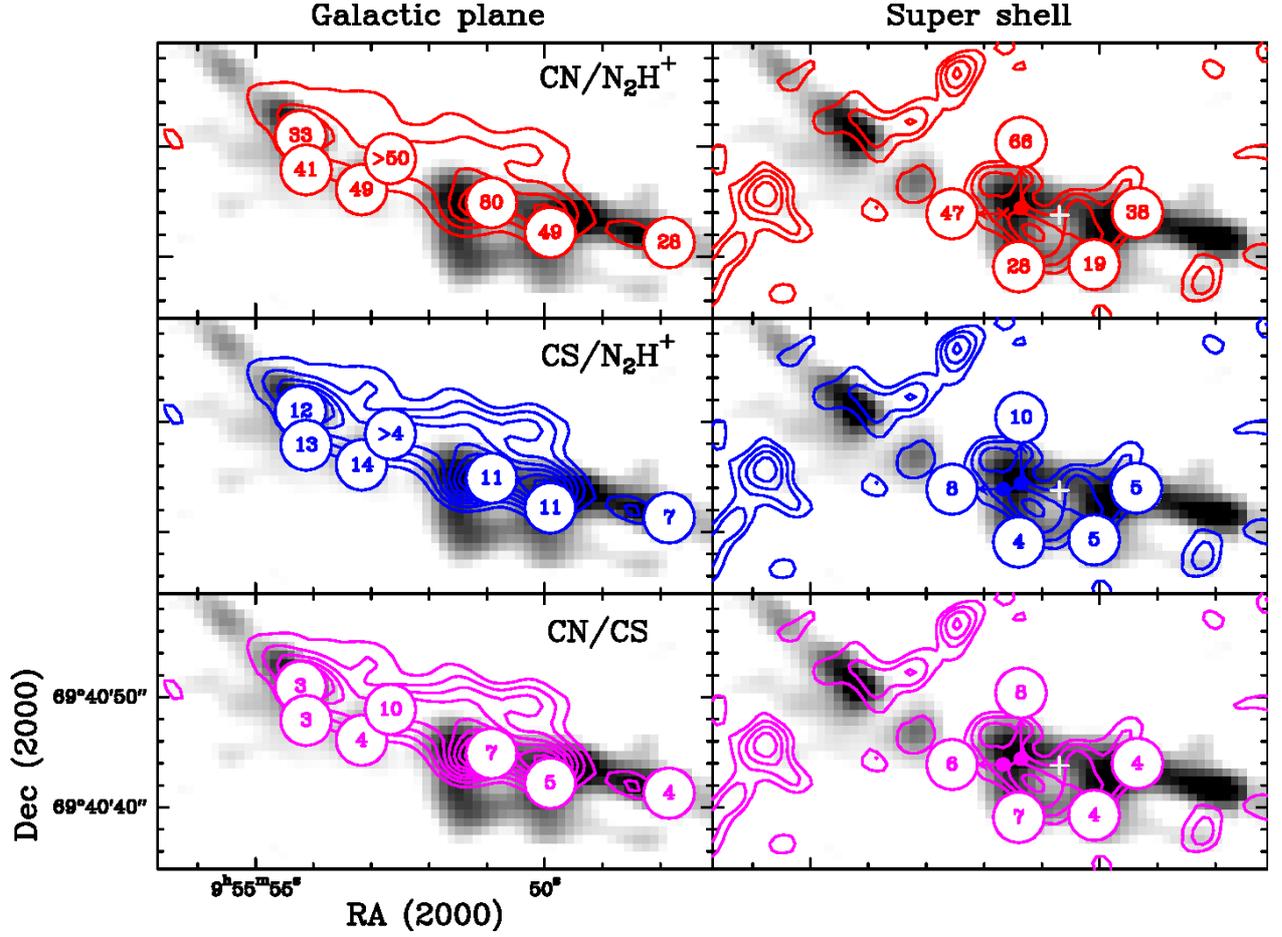}
 \caption{[CN]/[N$_2$H$^+$], [CS]/[N$_2$H$^+$] and [CN]/[CS] abundance ratios for the positions listed in Table 2 in the galacic plane (left) and
the supershell associated with SN~41.9+58 (right) superposed to the N$_2$H$^+$ 1$\rightarrow$0 integrated intensity map (grey scale).
For comparison, on the left column we plot the contours of the integrated intensity emission of the H (41) $\alpha$ line, and
on the right, the integrated intensity emission of the SiO 2$\rightarrow$1 line (Garc\'{\i}a-Burillo et al. 2001). The white cross indicates
the position of SN~41.9+58.}
 \label{fig9}
\end{figure*}

Significant variations, i.e. higher than the uncertainties, are found in the [CN]/[N$_2$H$^+$] ratio across the galaxy.
This ratio takes values $\approx$30 in GP1 and GP6 located in the outer $x1$ orbits and increases to $\approx$80 towards GP4.
Values $>$50 are found towards GP5 and SS1, both close to W2. 
The minimum value, [CN]/[N$_2$H$^+$]$\approx$19, is found in the southern part of the supershell.
In Fig.~\ref{fig9} we compare the obtained column density ratios
with the integrated intensity maps of the N$_2$H$^+$ 1$\rightarrow$0, H(41)$\alpha$ lines and SiO 2$\rightarrow$1 lines.
It is remarkable that there is a good correlation between the values of the [CN]/[N$_2$H$^+$] ratio 
and the emission of the  H(41)$\alpha$ line, supporting the interpretation that the variations of the [CN]/[N$_2$H$^+$] ratio 
are related to the variations of the local UV field. We do not find any correlation between the [CN]/[N$_2$H$^+$] ratio 
and the SiO emission.

The [CN]/[CS] ratio behaves similarly to [CN]/[N$_2$H$^+$]. It takes values $\approx$3.5 in GP1 and GP6 located in the outer $x1$ orbits 
and increases to $>$5 towards GP4, GP5, and SS1. In the case of CN1, we estimate a higher value of, 
[CN]/[CS]$\approx$10. There is a good correlation between the [CN]/[CS] ratio and the H(41)$\alpha$ emission.

Contrary to the [CN]/[N$_2$H$^+$] and [CN]/[CS] ratios, the [CS]/[N$_2$H$^+$] ratio remains quite constant across 
the galactic plane with values around
$\sim$12. There is no evidence of variation of this ratio with the local UV field. When comparing the galactic plane positions
with those of the supershell, we realized that the ratio tends to be a factor of 2 lower in the southern part of
the supershell, although always within the uncertainty of our calculations. 

We used the C$^{18}$O map from Weiss et al. (2001b) to determine the total molecular hydrogen column densities and
absolute fractional abundances. The C$^{18}$O column densities vary by only a factor of $\sim$2 among the different knots 
with the largest values in GP1 and GP6. 
Molecular abundances have been derived from the N(X)/N(C$^{18}$O) column density ratio 
assuming [C$^{18}$O]/[$^{12}$CO]=0.005 and [$^{12}$CO]=4$\times$10$^{-5}$, which are the average values
derived by Weiss et al. (2001b) on basis of their multitrasitional study of CO, $^{13}$CO and C$^{18}$O. 
Note that these values correspond to a $^{16}$O/$^{18}$O ratio of 200  which is different from the Milky Way 
standard value of $\sim$500$-$600. 
We are assuming a constant  C$^{18}$O abundance all over the galaxy which is
an approximation since the C$^{18}$O abundance is dependent on
the environment (UV flux, density, temperature).
However, we are interested in the dense molecular gas (n(H$_2$)$>$1$\times$10$^5$~cm$^{-3}$) and the
C$^{18}$O abundance is expected to remain roughly constant in this component. Following this procedure 
we derived the molecular abundances shown in
Table~2. The N$_2$H$^+$ abundance is $\sim$3$\times$10$^{-11}$ for all the positions, 
corroborating that N$_2$H$^+$ is a good tracer of the dense molecular gas. An abundance of a few 10$^{-10}$ is measured for 
CS. Small changes, less of a factor of 2, are found in the abundances of this species that are
within the uncertainties of our calculations. Only the CN abundance presents significant 
variations within the galaxy, being larger by a factor $\sim$3 in GP4 than in the outer posisions, GP1 and GP6,
and the southern supershell positions, SS2 and SS3. 

\begin{table*}
\caption{Selected regions for the chemical model$^1$.}
\label{tableClumps}
\begin{tabular}{cccccccccccc} \\ 
\hline
Point & \multicolumn{2}{c}{Offset} & Vel               &  N(C$^{18}$O)                     &   X(CN)                   & X(N$_2$H$^+$)  &   
X(o-C$_{3}$H$_2$)$^2$   &  X(CS)     &   [CN]/[N$_2$H$^+$]  &  [CS]/[N$_2$H$^+$] &  [CN]/[CS]  \\
      & \multicolumn{2}{c}{($\arcsec$)} & (km s$^{-1}$)  & ($\times$10$^{16}$ cm$^{-2}$)     &   ($\times$10$^{-9}$)     &   ($\times$10$^{-11}$) &
($\times$10$^{-10}$)                  &  ($\times$10$^{-10}$)    &    &   &       \\
\hline
 GP1 &  +12.1  &  +4.1 & 327.4  &  2.2$_{-1.0}^{+1.2}$  &  0.9$_{-0.3}^{+0.8}$    &   3.0$_{-1.3}^{+1.8}$  &  3.2$_{-1.7}^{+3.0}$    &  3.0$_{-1.1}^{+3.9}$  &
 33$_{-11}^{+17}$  &  12$_{-6}^{+8}$  &  3.0$_{-0.5}^{+0.5}$   \\ 
 GP2 &  +11.6  & +1.0  & 289.0  &  1.8$_{-0.8}^{+1.0}$  &  1.0$_{-0.3}^{+0.8}$    &   2.8$_{-1.4}^{+1.3}$  &  2.6$_{-1.4}^{+2.7}$    &  3.2$_{-1.3}^{+3.6}$  &
 41$_{-13}^{+22}$  &  13$_{-6}^{+10}$  & 3.3$_{-0.5}^{+0.5}$   \\ 
 GP3 &   +6.6  & -0.8  & 263.4  &  1.8$_{-0.8}^{+1.0}$  &  1.4$_{-0.5}^{+1.0}$    &   3.0$_{-1.6}^{+1.4}$  &  3.0$_{-1.5}^{+3.3}$    &  3.8$_{-1.5}^{+4.3}$  & 
 49$_{-16}^{+23}$  &  14$_{-6}^{+10}$  & 3.7$_{-0.7}^{+0.6}$  \\ 
 GP4 &  -5.1   & -2.0  &  97.0  &  0.9$_{-0.4}^{+0.5}$  &  3.0$_{-1.1}^{+2.1}$    &   4.3$_{-2.1}^{+2.0}$  &  2.6$_{-1.4}^{+2.8}$    &  3.5$_{-1.2}^{+5.2}$  &
 80$_{-28}^{+34}$  &  11$_{-5}^{+8}$  & 7.4$_{-1.5}^{+1.4}$ \\
 GP5 & -10.4   & -4.7  &  97.0  &  1.9$_{-0.9}^{+1.1}$  &  1.4$_{-0.4}^{+1.0}$    &   3.2$_{-1.6}^{+1.5}$  &  2.4$_{-1.2}^{+2.7}$    &  2.6$_{-1.0}^{+3.6}$  & 
49$_{-16}^{+23}$  &   11$_{-5}^{+7}$  & 4.8$_{-0.9}^{+0.9}$  \\
 GP6 & -21.1   & -5.6  &  84.2  &  2.1$_{-1.0}^{+1.1}$  &  0.8$_{-0.3}^{+0.6}$    &   3.1$_{-1.5}^{+1.5}$  &  1.5$_{-0.8}^{+1.7}$    &  1.8$_{-0.7}^{+2.5}$  &
28$_{-10}^{+13}$  &   7$_{-3}^{+6}$   &  3.8$_{-0.6}^{+0.6}$  \\
\\
 SS1 &  -1.2   & -2.0  & 173.8  &  1.0$_{-0.5}^{+0.5}$  &  2.0$_{-0.7}^{+1.6}$    &   4.8$_{-2.4}^{+2.3}$  &  3.1$_{-1.2}^{+3.5}$    &  2.9$_{-1.0}^{+4.3}$  &  
47$_{-16}^{+23}$  &   8$_{-5}^{+5}$   &  6.2$_{-1.2}^{+1.0}$   \\
 SS2 &  -2.6   & -7.8  & 161.0  &  0.6$_{-0.3}^{+0.3}$  &  0.9$_{-0.3}^{+0.8}$    &   3.8$_{-1.8}^{+1.9}$  &  1.4$_{-0.7}^{+1.6}$    &  1.1$_{-0.7}^{+1.8}$  &
28$_{-10}^{+13}$  &  4$_{-2}^{+3}$    &  7.5$_{-1.4}^{+1.4}$   \\
 SS3 &  -9.4   & -7.6  & 148.2  &  1.4$_{-0.7}^{+0.7}$  &  0.5$_{-0.2}^{+0.4}$    &   2.7$_{-1.3}^{+1.4}$  &  1.4$_{-0.7}^{+1.7}$    &  1.0$_{-0.4}^{+1.5}$  &   
19$_{-7}^{+11}$   &  5$_{-3}^{+4}$    &  4.2$_{-0.8}^{+0.6}$   \\
 SS4 &  -13.3  & -2.9  & 148.2  &  1.5$_{-0.7}^{+0.7}$  &  0.9$_{-0.3}^{+0.7}$    &   2.7$_{-1.3}^{+1.3}$  &  2.2$_{-1.0}^{+2.1}$    &  2.4$_{-1.1}^{+2.6}$  & 
38$_{-13}^{+19}$  &   5$_{-3}^{+4}$    & 4.0$_{-0.7}^{+0.7}$  \\   
 SS5 &  -2.8   & -2.5  & 135.4  &  1.1$_{-0.5}^{+0.5}$  &  2.4$_{-0.8}^{+1.9}$    &   4.2$_{-2.0}^{+2.0}$  &  2.6$_{-1.3}^{+2.9}$    &  2.6$_{-0.9}^{+3.9}$  &  
66$_{-23}^{+29}$  &  10$_{-5}^{+7}$    & 8.3$_{-1.7}^{+1.6}$  \\   
\\
 CN1 &  +4.0   &  +2.0 & 327.4  &  0.3$_{-0.1}^{+0.1}$  &  2.0$_{-0.7}^{+1.8}$    &   $<$4.0$^*$           &  $<$3.0$^*$             &  1.8$_{-0.6}^{+3.1}$  &
$>$50             &  $>$4              & 9.8$_{-2.0}^{+2.0}$  \\
\hline
\end{tabular}

\noindent
$^1$Relative abundaces wrt. H$_2$; $^2$ Contaminated with methanol (see text); $^*$ 3$\sigma$ limit. 

\end{table*}

\subsection{Other species: H$_2$CO, HC$_3$N and CH$_3$CN}

In order to have a more comprehensive view of the chemistry in M~82 we have compared our new data with
previous interferometric images. For this aim, we degraded the 
angular resolution of all our images to $\approx$5.9$"$ ($\approx$111~pc), which is the lowest from the data set shown 
in Table 1, and used the total velocity integrated line intensity in our calculations. The Gaussian fit parameters
towards E1, E2, W2 and W1 are shown in Table A.1. 
 
Column densities and column density ratios were calculated following the same procedure as in
Sect.~4.1. In Table~3, we show the values obtained for E1, E2, W2 and W1. 
The [CS]/[H$^{13}$CO$^+$] and [H$^{13}$CO$^+$]/[C$^{18}$O] are quite uniform across the galaxy. This suggests that 
CS, H$^{13}$CO$^+$ and C$^{18}$O are good tracers of dense gas in a wide range of physical conditions. 
According to these results, E1 and W1 present the highest total molecular hydrogen column
densities. This is consistent with the detection of complex molecules like NH$_3$, HC$_3$N, CH$_3$CN and CH$_3$OH
towards them (Weiss et al. 2001a, Mart\'{\i}n et al. 2006, Aladro et al. 2011). 
The small hydrocarbons c-C$_3$H$_2$ and C$_4$H are also good tracers of
low and high UV PDRs (Fuente et al. 2003, Pety et al. 2005, Pilleri et al. 2013, Cuadrado et al. 2014).
Taking into account that the abundance of c-C$_3$H$_2$ is very likely overestimated by a factor of $\sim$2 in E1 and W1
(see Sect. 4), we conclude that the abundance of c-C$_3$H$_2$ could be
enhanced by a factor of $\sim$3 in the inner $x2$ orbits, in agreement with the trend observed in CN.
Moreover, the derived H$_2$CO abundance is similar to that derived by Guzm\'an et al. (2013, 2014) in the 
Horsehead nebula and consistent with the interpretation of the H$_2$CO emission coming from PDRs. 
Summarizing, the overall chemical behavior observed in M~82 is well interpreted as the consequence
of the effect of UV radiation on the molecular chemistry. This produces an increase in the abundance of PDR 
tracers (CN, c-C$_3$H$_2$, H$_2$CO)
towards E2 and W2, following the spatial distribution of HII regions. The detection of complex molecules shows that
a fraction of the gas is protected from the UV radiation in the interior of large molecular clouds. The amount of gas
in this shielded component is higher in the outer $x1$ bar orbits.  

We remind that the position W2 is related with the supernova explosion SN~41.9+58 that is launching ionized gas out of the galaxy plane. 
The detection of SiO by Garc\'{\i}a-Burillo et al. (2001) proved the existence of shocks
and their effect on the gas chemistry. 
H$_2$CO, CH$_3$OH and CS are considered good tracers of shocks in galactic and extra-galactic
environments (see e.g. Bachiller et al. 2001, Garc\'{\i}a-Burillo et al. 2000, Usero et al. 2006). However, our data do not present any evidence of 
CS abundance enhancement because of possible shocks. In fact, the [CS]/[N$_2$H$^+$] ratio is slightly lower in the super-shell.
This suggests that only a small fraction of the gas is affected by shocks, which is
also consistent with the low average SiO abundance determined by Garc\'{\i}a-Burillo et al. (2001), X(SiO)$\sim$1$\times$10$^{-10}$,
when compared with the SiO abundances, $\sim$10$^{-8}$$-$10$^{-6}$, found in the shocks associated with galactic star forming regions (Mart\'{\i}n-Pintado
et al. 1992, Bachiller et al. 2001). In the case of H$_2$CO anc CH$_3$OH we cannot do a detailed analysis because their lines are blended with
other species.

\begin{table}
\caption{Fractional abudances and abundance ratios.}
\centering    
\begin{tabular}{lcccc  } 
\hline 
\multicolumn{1}{l}{} & \multicolumn{1}{c}{E1}& 
\multicolumn{1}{c}{E2} &  \multicolumn{1}{c}{W2}  & 
\multicolumn{1}{c}{W1}    \\ 
\\ \hline
N(C$^{18}$O) ($\times$ 10$^{16}$)                      & 10$_{-4}^{+6}$     &  4.1$_{-1.7}^{+2.5}$  &  4.8$_{-2.0}^{+2.7}$  & 11$_{-4}^{+7}$   \\
\\ 
X(CN)  ($\times$ 10$^{-9}$)                             & 1.4$_{-0.4}^{+0.5}$  &  3.2$_{-0.8}^{+1.0}$     &  4.0$_{-1.0}^{+1.4}$   &  1.2$_{-0.3}^{+3.8}$    \\
X(N$_2$H$^+$)   ($\times$ 10$^{-11}$)                   & 2.4$_{-0.6}^{+0.6}$  &  3.1$_{-0.4}^{+0.6}$  &  4.4$_{-0.9}^{+0.8}$  &  2.0$_{-0.4}^{+0.4}$  \\
X(CS)  ($\times$ 10$^{-10}$)                            & 5.2$_{-1.8}^{+2.4}$  &  7.2$_{-2.5}^{+3.6}$  &  7.8$_{-2.8}^{+3.6}$  &  4.3$_{-1.4}^{+2.0}$   \\
X(H$^{13}$CO$^+$) ($\times$ 10$^{-11}$)                 & 1.8$_{-0.7}^{+0.7}$  &  1.3$_{-0.6}^{+0.6}$  &  2.4$_{-0.8}^{+1.0}$  &  1.3$_{-0.5}^{+0.5}$  \\
X(o-C$_3$H$_2$)    ($\times$ 10$^{-10}$)                & 2.0$_{-1.2}^{+2.0}$  &  2.8$_{-1.8}^{+2.7}$  &  3.4$_{-2.2}^{+3.0}$  &  2.4$_{-1.6}^{+2.4}$   \\
X(p-H$_2$CO)     ($\times$ 10$^{-11}$)                  & 7.4$_{-3.6}^{+5.8}$  &  $>$9.6$_{-4.8}^{+7.8}$$^*$  &  $>$5.6$_{-2.8}^{+4.4}$$^*$  & 4.6$_{-2.2}^{+3.8}$  \\
X(HC$_3$N)       ($\times$ 10$^{-11}$)                  & 8.1$_{-6.5}^{+12}$   &  $<$13$_{-11}^{+19}$$^* $    &  $<$31$_{-25}^{+46}$$^*$   &  5.3$_{-4.2}^{+8.8}$      \\
X(CH$_3$CN)    ($\times$ 10$^{-11}$)                    & 1.4$_{-0.6}^{+2.0}$  &  $<$1.0$^{**}$            &   $<$0.8$^{**}$       &   1.0$_{-0.4}^{+1.3}$     \\
\\
$[CN]/[N_2H^+]$     &   56$_{-7}^{+13}$        &  101$_{-16}^{+29}$     & 92$_{-14}^{+18}$      &  58$_{-8}^{+14}$  \\
$[CN]/[CS]$         &   2.8$_{-1.0}^{+1.8}$    &  4.7$_{-1.7}^{+1.5}$   &  5.6$_{-2.1}^{+1.6}$  &  2.9$_{-0.9}^{+0.9}$ \\
$[CS]/[N_2H^+]$     &   20$_{-6}^{+6}$         &  17 $_{-8}^{+13}$      &  23$_{-8}^{+9}$       &  21$_{-7}^{+7}$  \\
$[H^{13}CO^+]/[N_2H^+]$  & 0.7$_{-0.2}^{0.2}$  &  0.5$_{-0.2}^{+0.2}$ & 0.5$_{-0.2}^{+0.2}$     &  0.6$_{-0.2}^{+0.}$  \\
\hline
\end{tabular}
\tablefoot{Beam (5.9$\arcsec$) average column densities resulting from LVG calculations in a grid with 
T$_k$=50--150 K, and n(H$_2$)=1$\times$10$^5$, 5$\times$ 10$^5$. Errors correspond to the minimum and maximum values in the grid.
$^*$ Reasonable guess since the HC$_3$N and H$_2$CO are blended at this position.$^{**}$ Upper limit have been derived assuming $\Delta$v=50~km~s$^{-1}$,
T$_k$=50~K and n(H$_2$)=5$\times$10$^5$~cm$^{-3}$.}
\label{tableOffsetsColumnDensities}
\end{table}

\section{Chemical model}
We used the Meudon PDR code 1.4.4 (Le Petit et al. 2006) to model the chemistry
in M~82. In our calculations, we assume that the ISM is composed of clouds
bathed by an intense UV field. We simulate each cloud by a uniform plane-parallel layer illuminated from the two sides. 
This layer has constant density (n=n(H)+2$\times$n(H$_2$)=4$\times$10$^5$~cm$^{-3}$) and the gas kinetic temperature is 
calculated by detailed heating and cooling balance. The adopted initial elemental abundances are the same as in Fuente et al. (2008). 
We run a grid of models varying the cloud size, UV field and cosmic rays ionization rate.

\begin{figure*} 
 \centering
 \includegraphics[width=0.9\textwidth] {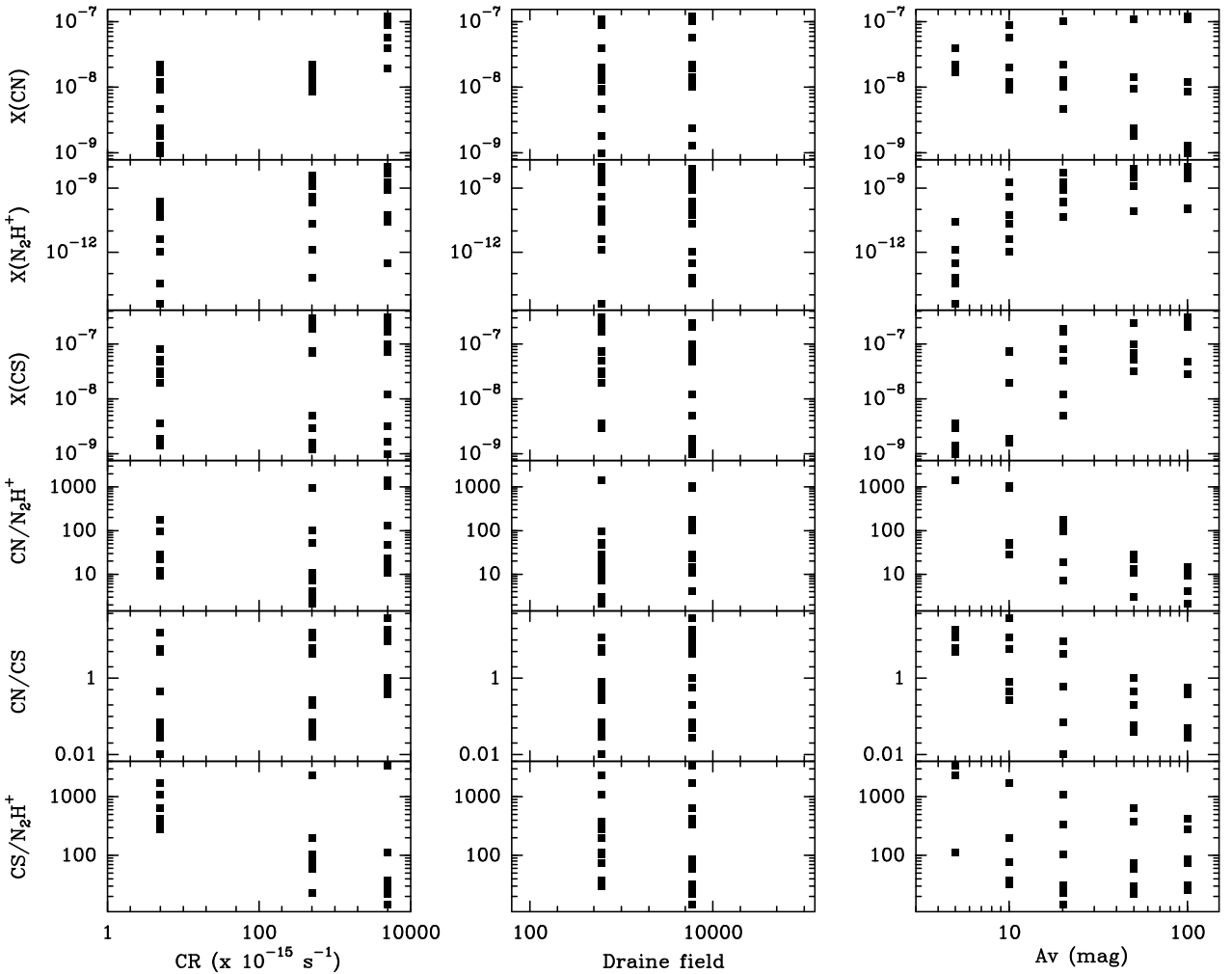}
 \caption{Model predictions as a function of the cosmic ray ionization rate, Draine field and cloud size. Model parameters and results 
are shown in Table~4.}
 \label{fig10}
\end{figure*}

Our grid includes two values of the Draine field, 6$\times$10$^3$ and  6$\times$10$^2$. The first value
was determined by Fuente et al. (2005, 2006, 2008) from the CO$^+$, HCO$^+$ and HOC$^+$ observations. The second value is
considered to account for the possibility of regions with a lower UV field within the galaxy. 
To investigate the effect of the enhanced cosmic-ray flux on the chemistry, we have repeated our model calculations with
$\zeta$=5$\times$10$^{-17}$ and 5$\times$10$^{-15}$~s$^{-1}$. The first value corresponds to the cosmic rays ionization rate
in the molecular clouds of our Galaxy (Indriolo et al. 2014). The second is the value estimated by Suchkov et al. (1993) to account for the physical
conditions of the molecular gas in M~82.
We are aware that X-rays are intense and could also play an important role in the chemistry (Fuente et al. 2008, 
Spaans \& Meijerink 2007).
Our code does not include the X-rays in the chemical calculations. We investigate the effect that X-rays could have
on the chemistry by increasing the cosmic rays ionization rate to 5$\times$10$^{-14}$~s$^{-1}$. To increase 
the cosmic rays ionization rate reproduces reasonably well the effect of X-rays on the molecular chemistry although cannot 
account for the effect on the molecular gas heating. Finally, in our simple model, the cloud size is given by the total 
cloud visual extinction.
We considered five values: 5~mag, 10~mag, 20~mag, 50~mag and 100~mag.

In Fig.~\ref{fig10} and Table 4 we show the cloud average molecular abundances as a function of the different model parameters. The large
dispersion in the values is not unexpected since all the parameters are varied by several orders of magnitude.
The cloud average abundances of CN, N$_2$H$^+$, CS and HCO$^+$ increase with the cosmic rays ionization rate.
This increase is especially important in the case of N$_2$H$^+$. While the abundances of the other species increase by a factor of
a few, the abundance of N$_2$H$^+$ increases by two orders of magnitude between
$\zeta$=5$\times$10$^{-17}$ and 5$\times$10$^{-14}$ s$^{-1}$. Surprisingly, the N$_2$H$^+$ abundance is quite constant in the
studied knots. This suggests that the chemical changes we observe are not related with variation in the X-rays flux. 
We cannot discard, however, that
variations in the UV field and cloud sizes could balance to keep the N$_2$H$^+$ abundance constant in spite of 
the variation of the X-rays flux (Strickland \& Heckman 2007).

The cloud size is the parameter that has the largest impact on the results of our model. N$_2$H$^+$ has a negligible 
abundance in clouds with A$_v$$<$10~mag and abundances varying between 10$^{-10}$ and 10$^{-9}$ for larger clouds. 
This corroborates our assumption that this species is an excellent tracer 
of the UV-protected molecular gas. 
The [CN]/[N$_2$H$^+$] ratio is a good measure of the cloud size in the case of one single cloud component. 
Fuente et al. (2008) proposed the existence of two cloud components towards E1: most of the mass, $\sim$87\%, is locked in small 
clouds of A$_v$$\sim$5 mag and the rest is forming A$_v$$>$50~mag clouds. In this case, the [CN]/[N$_2$H$^+$] ratio would depend
on both, the fraction of each cloud component and the assumed size for the large clouds.

The cloud average CS abundance is quite constant for sizes between $\sim$5 and $\sim$20 mag but increases by almost one
order of magnitude for larger sizes. The large values of the CS abundance in the 50~mag
clouds are not reliable. The CS abundance is very dependent on the assumed elemental sulphur abundance.
Our gas phase models does not consider the adsorption of molecules on the grain surfaces. Within our galaxy, 
the sulphur abundance measured in PDRs is a factor $>$4 lower than the solar value, suggesting that significant
sulphur depletion occurs even in highly irradiated environments (see Goicoechea et al. 2006). Moreover, our model assumes 
the solar value for the sulphur elemental 
abundance, 1.8$\times$10$^{-5}$, that could greatly differ from that in M~82 (Umeda et al. 2002, Origlia et al. 2004). 
Therefore, we do not consider that the values of the CS abundance predicted by our model are reliable.

The [CN]/[N$_2$H$^+$] ratio is a sensitive tracer of the cloud size for sizes $<$50~mag. 
Other parameters like the incident UV field and the cosmic rays ionization rate could also affect the 
predicted [CN]/[N$_2$H$^+$] ratio. As expected, the [CN]/[N$_2$H$^+$] ratio increases for higher UV fluxes, especially for
small clouds. On the contrary, the [CN]/[N$_2$H$^+$] ratio decreases with the cosmic rays ionization rate because of 
the increase of the N$_2$H$^+$ abundance. In the following, we discuss the cloud distribution in M~82 on basis 
of the [CN]/[N$_2$H$^+$] ratio. We use our a priori knowledge of the galaxy to constrain the
chemical models. We discuss the results for one- and two-component models.

\begin{figure} 
 \centering
 \includegraphics[width=0.5\textwidth] {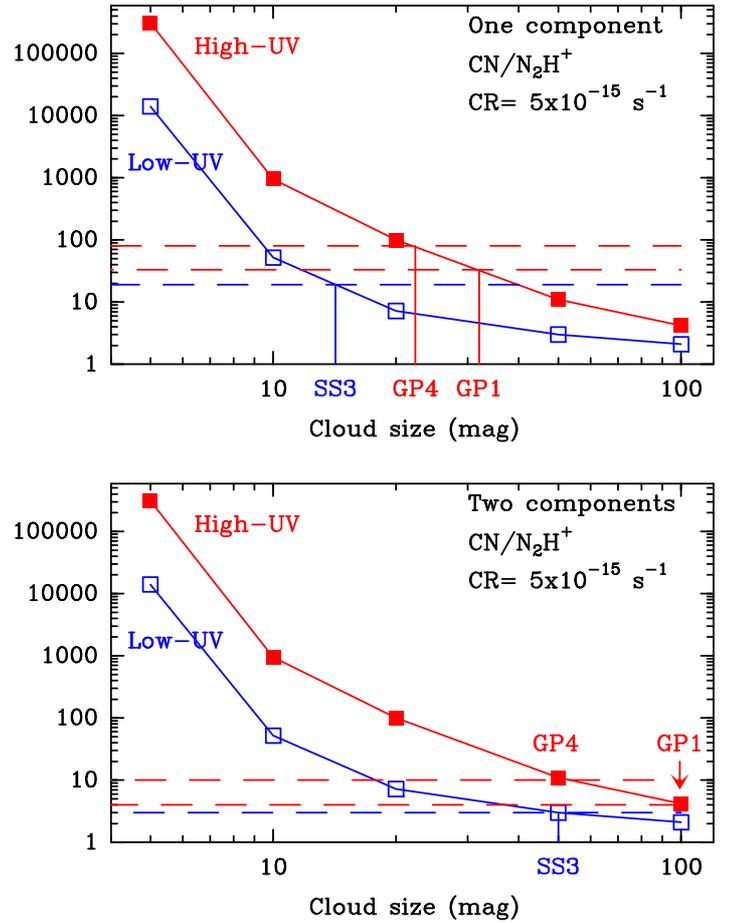}
 \caption{[CN]/[N$_2$H$^+$] abundance ratio as a function of the cloud size for models with CR=5$\times$10$^{-15}$ s$^{-1}$ and Draine fields of
6$\times$10$^3$ (red line) and $\times$10$^3$ (blue line). {\bf Top} Cloud sizes in the one-component model. {\bf Bottom} Sizes of the large cloud
component in the two-component model (see text). The values and errors for GP1, GP4 and SS3 are indicated.}
 \label{fig11}
\end{figure}

In Fig.~\ref{fig11}, we plot the [CN]/[N$_2$H$^+$] abundance ratio as a function of the cloud size for Draine fields
of 6$\times$10$^3$ (high-UV) and 6$\times$10$^2$ (low-UV). We adopt the value derived by Suchkov et al. (1993) 
for the cosmic rays ionization rate because it is the most likely. GP4 is the position with the highest value of 
the [CN]/[N$_2$H$^+$] ratio and also the emission peak of
the H(41)$\alpha$ line. It is reasonable to think that high-UV models, in red lines in  Fig.~\ref{fig11}, are more adequate 
to account for the molecular abundances at this position. One single cloud component with sizes of $\sim$22 mag could explain 
the observed [CN]/[N$_2$H$^+$] ratio of $\sim$80. 
Lower [CN]/[N$_2$H$^+$] ratios, $\sim$30, are measured towards the positions GP1 and GP6. Although less intense, these knots are also associated
with peaks in the H(41)$\alpha$ line emission which proves the presence of energetic UV photons
capable to ionize the atomic hydrogen (see Fig.~\ref{figa3}). Assuming the high-UV case, the lower [CN]/[N$_2$H$^+$] ratio can be explained by 
the existence of a population of larger clouds, $\sim$30~mag (see Fig.~\ref{fig11}). The existence of larger clouds is also consistent 
with the lower kinetic temperatures and higher densities measured by Weiss et al. (2001b) and the detection of complex molecules
at these positions.
Low values of the [CN]/[N$_2$H$^+$] ratio, $\sim$20$-$30, are also observed towards the SS2 and SS3 points placed in the 
molecular supershell. Since there is a lack of H(41) $\alpha$ emission towards these positions, we used the low-UV field models 
to interpret the [CN]/[N$_2$H$^+$] ratio and obtained that the molecular emission could come from translucent clouds of $\sim$14 mag. 

The variation of the average cloud size we conclude from our one-component model is compatible with
the results of Fuente et al. (2008) who interpreted the abundances in E1 in a two-component scenario.
We can use a two-component model to explain our observations. In this case, the CN emission would come from
the small (with Av$\sim$5~mag) and large (with Av$\sim$50~mag) molecular clouds while the
N$_2$H$^+$ emission would come only from large molecular clouds.  
Fuente et al. (2008) estimated that only $\sim$13\% of the molecular gas is forming large molecular clouds in E1.
Assuming the same fraction at all positions, the [CN]/[N$_2$H$^+$] ratio in the large cloud component would be 
about 0.13 times the observed one, i.e. $\sim$10 in GP4, $\sim$4 in GP1 and $\sim$3 in SS3, i.e., we would have clouds of $>$50~mag
in the three positions, with the largest ones, $\sim$100 mag, towards GP1. 
Summarizing, both, the one and two-components models, come to the conclusion that the highest concentration of
UV-protected molecular gas is found towards the $x1$ orbits. The detection of complex molecules
is better understood in terms of two populations of clouds with a small fraction of the gas, $\sim$13\%,
locked in giant molecular clouds where the gas is UV protected and can form large molecules.

Of course, the two models are very simple and have to be understood as a guide to interpret
the molecular chemistry in this starburst galaxy. As commented above, our model neglects the surface chemistry and 
X-rays effets that could be important for some species (see discussion above). Moreover our gas phase model assumes 
steady-state chemistry. 
The characteristic time for PDR chemistry is $\tau$$\sim$1/($\kappa_d$ G$_0$), where $\kappa_d$ is the photocissociation 
rate in s$^{-1}$ and G$_0$ is the incident UV field in Habing units. The value of $\kappa_d$ depends on the 
extinction from the illuminating source as $\kappa_d$$\sim$$\kappa_0$$\times$ exp($-$b~A$_v$). 
Assuming a typical value of $\kappa_0$=10$^{-9}$~s$^{-1}$, b=1.8 and the incident mean UV
interstellar field in M~82, G$_0$=10$^4$ Habing fields, the characteristic time is $>$10~Myr for A$_v$$>$2.5~mag. 
This means that if the last starburst episode took place about $\sim$5$-$10~Myr ago, 
the chemistry might be out of equilibrium in the clouds interior. Our steady-state scenario is therefore only
adequate for the cloud surfaces and small clouds. The shortcoming of time-dependent models is that
depend on the initial conditions and require of a more sophisticated analysis that is beyond the scope of this paper.
Subsequent improvement of our current interpretation would require to couple the star formation history in
this galaxy with a time-dependent PDR code including surface chemistry
(Bayet et al. 2009, Viti et al. 2014, Bisbas et al. 2012).

For all species, the estimated fractional abundances in M~82 are lower than those predicted by our gas-phase model.
Absolute fractional abundances are difficult to compare with models because of the large uncertainty in the total
molecular gas column density. Weiss et al. (2001b) derived the total molecular gas column densities in M~82 using 
different methods: LVG calculations, Local Thermodynamic Equilibrium (LTE) solution and the standard X$_{CO}$ conversion factor,
and their results agree within a factor $\sim$3-4. In this paper, we derived the molecular column density towards the
different points using the C$^{18}$O data and assuming the CO abundance derived by Weiss et al. (2001b), X(CO)=5$\times$10$^{-5}$.
We would like to note that this value is already lower than the value predicted by our model for large clouds in which 
essentially all the carbon is in CO with an abundance of $\sim$1.3$\times$10$^{-4}$. Therefore our fractional abundances
are accurate within a factor of 4--10 and consequently, roughly consistent with model predictions.

\begin{table*}
\caption{Chemical models$^1$}
\label{tableClumpsColumnDensityRatios}
\begin{center}    
\begin{tabular}{cccccccccc} 
\hline
Model & CR ($\times$10$^{-17}$) &   Cloud size (mag)    &  Draine field  &  X(CN)  & X(N$_2$H$^+$) & X(CS) & [CN]/[N$_{2}$H$^{+}$] & [CN]/[CS]  & [CS]/[N$_2$H$^+$] \\
\hline
\multicolumn{10}{c}{High-UV} \\
1 & 5000  & 100  & 6e3   &  1.2e-7   &   8.0e-9   &    2.1e-7   &   15   &   0.56   &   27 \\ 
2 & 500   & 100  & 6e3   &  1.2e-8   &   2.8e-9    &    2.4e-7   &   4.2   &   0.05   &   85 \\ 
3 & 5     & 100  & 6e3   &  1.3e-9   &   1.1e-10   &    4.7e-8   &   12    &   0.03  &   439 \\ 
\\
4 & 5000  &  50   &  6e3  &  1.1e-7 &  4.6e-9   &  1.0e-7  &   23  &   1.0  &   22  \\  
5 & 500   &  50   &  6e3  &  1.4e-8 &  1.2e-9   &  7.0e-8  &   11  &   0.2  &   58 \\
6 & 5     &  50   &  6e3  &  2.4e-9 &  8.1e-11  &  5.2e-8  &   29  &  0.05  &   641 \\  
\\
7 & 5000   &  20  &   6e3 &  1.0e-7 &  8.3e-10  &  1.2e-8  &  130  &   9.0   &  14  \\
8 & 500    &  20  &   6e3 &  2.2e-8 &  2.2e-10  &  5.0e-9  &  100  &   4.4   &  22  \\
9 & 5      &  20  &   6e3 &  1.0e-8 &  2.4e-10  &  8.0e-8  &  2173 &   0.07  &  333 \\
\\
10 & 5000   &  10  &  6e3  &  5.7e-8 &  5.3e-11  &  1.7e-9  &  1055    &  38  &  32  \\ 
11 & 500    &  10  &  6e3  &  2.0e-8 &  2.1e-11  &  1.6e-9  &   952    &  12  &  76  \\
12 & 5      &  10  &  6e3  &  1.2e-8 &  1.1e-12  &  1.9e-9  &  1.0e4   &  6   &  1727  \\
\\
13 & 5000   &  5   &  6e3  &  1.9e-8  &  3.0e-13 &  1.0e-9  &  6.4e4   &  19   &  3333 \\        
14 & 500    &  5   &  6e3  &  2.0e-8  &  6.4e-14 &  1.2e-9  &  3.1e5   &  16   &  18750 \\ 
15 & 5      &  5   &  6e3  &  2.2e-8  &  3.6e-14 &  1.4e-9  &  6.1e5   &  16   &  38889 \\
\\ \hline
\multicolumn{10}{c}{Low-UV}\\
16 & 5000  & 100  & 6e2   & 1.1e-7   &   1.0e-8   &   3.1e-7   &  11    &  0.38   &   30 \\
17 & 500   & 100  & 6e2   & 8.6e-9   &   4.0e-9  &   3.0e-7   &  2.1   &  0.03  &   74 \\
18 & 5     & 100  & 6e2   & 9.9e-10  &   1.0e-10  &   2.8e-8   &  9.4   &  0.03  &   270 \\
\\                              
19 & 5000   &  50  &  6e2  &   1.1e-7  &   8.2e-9  &  2.4e-7  &  13   &   0.45  &  29 \\  
20 & 500    &  50  &  6e2  &   9.4e-9  &   3.2e-9  &  2.4e-7  &   3   &   0.04  &  75 \\ 
21 & 5      &  50  &  6e2  &   1.8e-9  &   8.4e-11 &  3.2e-8  &  122  &   0.06  &  380 \\    
\\
22 & 5000   &  20  &  6e2  &   1.0e-7  &   5.5e-9  &   1.7e-7  &   19  &  0.59  &  31 \\             
23 & 500    &  20  &  6e2  &   1.3e-8  &   1.8e-9  &   1.9e-7  &   7   &  0.07  &  105 \\
24 & 5      &  20  &  6e2  &   4.6e-9  &   4.6e-11 &   5.0e-8  &   98  &  0.01  &  1086 \\
\\
25 & 5000   &  10  &  6e2  &  8.9e-8   &  1.9e-9   &  7.2e-8  &   47  &   0.81  &  37 \\   
26 & 500    &  10  &  6e2  &  2.0e-8   &  3.8e-10  &  7.6e-8  &   52  &   0.26  &  200 \\
27 & 5      &  10  &  6e2  &  9.0e-9   &  4.3e-12  &  2.0e-8  &   28  &   0.45  &  4651 \\
\\
28 & 5000   &  5   &  6e2  &  4.0e-8   &   2.8e-11 &   3.2e-9  &  1428  &   12  &  114  \\  
29 & 500    &  5   &  6e2  &  1.8e-8   &   1.3e-12 &   3.0e-9  &  1.4e4 &    6  &  2307 \\   
30 & 5      &  5   &  6e2  &  1.7e-8   &   4.0e-15 &   3.6e-9  &  4.3e6 &    5  &  9E5 \\ 
 \hline
\end{tabular}
\end{center}
$^1$ Notation: 6e3 = 6$\times$10$^3$
\end{table*}

\section{PDR ans shock chemistry in M~82}
M~82 has been subject of succesive starburst episodes in the last $\sim$200 Gyr. 
The most recent ($\sim$5~Myr) took place throughout the central regions of M~82 and was particularly intense
in the $x2$ orbits and along the stellar bar. The negative feedback effects from this starburst activity
is determining the existence and duration of future starburst episodes. We have carried out a 
chemical study of the molecular gas at scales of $\sim$100 pc in order to investigate
the feedback effects and the future star formation activity of the galaxy.

Our main result is that the chemistry of the molecular gas is determined by the intense UV radiation
produced by the massive stars. There is a systematic variation of the studied molecular abundance ratios 
with the intensity of the H(41)$\alpha$ emission that corroborates that
the whole nucleus ($\sim$650~pc), down to scales of $\sim$100~pc, behaves like a giant PDR. Within
this PDR, we have chemical variations due to changes in the local UV field and in the structure
of the molecular clouds. There are three well differentiated regions: 

\begin{itemize}
\item The inner $x2$ bar orbits 
that are associated with the most recent starburst and where the PDR tracers present their peak 
abundances. The detection of N$_2$H$^+$ in these regions proves that molecular clouds of 
$>$20~mag are present and therefore, the molecular gas reservoir to form new stars is
not exhausted.
  
\item The outer $x1$ bar orbits where most the UV-protected molecular gas is located. 
The detection of H(41)$\alpha$ shows that massive stars have already been formed in these clouds.

\item The molecular supershell associated with the supernova remnant SN~41.9+58. 
There are no signs of massive star formation activity in the southern part of this
super-shell although there are intense N$_2$H$^+$ knots that could
be sites for future star formation.  

\end{itemize}

The detection of a $\sim$500 pc molecular gas chimney and 
a super-shell in SiO indicates the occurrence of large-scale shocks in the disk-halo
interface (Garc\'{\i}a-Burillo et al. 2001).
We remind that the supershell is related with the supernova explosion SN~41.9+58 that is launching ionized gas out of the galaxy plane.  
However, our data do not present any evidence of possible shock chemistry. 
This suggests that only a small fraction of the dense molecular gas is affected by shocks, which is
also consistent with the low SiO abundance determined by Garc\'{\i}a-Burillo et al. (2001), X(SiO)$\sim$1$\times$10$^{-10}$,
when compared with the SiO abundances, $\sim$10$^{-8}$$-$10$^{-6}$, found in the shocks associated with galactic star 
forming regions (Mart\'{\i}n-Pintado et al. 1992, Bachiller et al. 2001). Higher spatial resolution observations
are required to detect the layer of molecular gas around the supershell whose chemistry is dominated by
shocks. This is also consistent with the moderate fraction of dense molecular gas ($\sim$2\%) that Salas et al.(2014) detected 
in the halo, in contrast with $\sim$25\% found by Walter et al. (2002) on basis of CO observations. 
Most of the expelled gas presents moderate densities and is not detected in dense molecular tracers. 

Summarizing, we present a comprehensive chemical study of the molecular gas in
the starburst galaxy M~82 using high spatial resolution (60--100~pc) 
interferometric images. Our chemical study shows that the feedback effects are strong in this evolved starburst.
In fact, the whole nucleus presents a PDR-like chemistry which suggests that UV radiation is driven by the
subsequent evolution of the ISM. Even though, the detection of N$_2$H$^+$ proves the existence of dense 
molecular gas enclosed in clouds of A$_v$$>$20~mag that could form new stars. 
The main reservoir of the dense molecular gas is located in the outer $x1$ orbits that might be the 
preferred site for a new generation of stars proceeding with the inside-outside scenario.

\begin{acknowledgements}
AF thanks the Spanish MINECO for funding support from grants CSD2009-00038 and AYA2012-32032.
SGB acknowledges support from Spanish grants CSD2009-00038, AYA2010-15169, AYA2012-32295,  
AYA2013-42227-P and from the Junta de Andalucia through TIC-114 and the  
Excellence Project P08-TIC-03531. PP acknowledges financial support from 
the Centre National d'Etudes Spatiales (CNES).
\end{acknowledgements}


\begin{thebibliography}{}

\bibitem[Achtermann 
\& Lacy(1995)]{1995ApJ...439..163A} Achtermann, J.~M., \& Lacy, J.~H.\ 1995, \apj, 439, 163 

\bibitem[Aladro et 
al.(2011)]{2011A&A...535A..84A} Aladro, R., Mart{\'{\i}}n, S., Mart{\'{\i}}n-Pintado, J., et al.\ 2011, \aap, 535, AA84 


\bibitem[Bachiller et 
al.(1997)]{1997A&A...319..235B} Bachiller, R., Fuente, A., Bujarrabal, V., et al.\ 1997, \aap, 319, 235 

\bibitem[Bachiller et 
al.(2001)]{2001A&A...372..899B} Bachiller, R., P{\'e}rez Guti{\'e}rrez, M., Kumar, M.~S.~N., \& Tafalla, M.\ 2001, \aap, 372, 899 

\bibitem[Bayet et al.(2009)]{2009ApJ...696.1466B} Bayet, E., Viti, S., 
Williams, D.~A., Rawlings, J.~M.~C., \& Bell, T.\ 2009, \apj, 696, 1466 

\bibitem[Bisbas et al.(2012)]{2012MNRAS.427.2100B} Bisbas, T.~G., Bell, 
T.~A., Viti, S., Yates, J., \& Barlow, M.~J.\ 2012, \mnras, 427, 2100 

\bibitem[Boger 
\& Sternberg(2005)]{2005ApJ...632..302B} Boger, G.~I., \& Sternberg, A.\ 2005, \apj, 632, 302 

\bibitem[Bregman et al.(1995)]{1995ApJ...439..155B} Bregman, J.~N., 
Schulman, E., \& Tomisaka, K.\ 1995, \apj, 439, 155 

\bibitem[Cernicharo (2012)]{cer12}Cernicharo, J. 2012, in Proceedings of the European Conference 
on Laboratory Astrophysics, Eur. Astron. Soc. Publ. Ser, eds. C. Stehlé, C. Joblin, \&
L. d’Hendecourt

\bibitem[Cuadrado et al.(2014)]{2014arXiv1412.0417C} Cuadrado, S., 
Goicoechea, J.~R., Pilleri, P., et al.\ 2014, arXiv:1412.0417 


\bibitem[F{\"o}rster Schreiber et al.(2003)]{2003ApJ...599..193F} 
F{\"o}rster Schreiber, N.~M., Genzel, R., Lutz, D., 
\& Sternberg, A.\ 2003, \apj, 599, 193 

\bibitem[Fuente et 
al.(1993)]{1993A&A...276..473F} Fuente, A., Mart\'{\i}n-Pintado, J., Cernicharo, J., \& Bachiller, R.\ 1993, \aap, 276, 473 

\bibitem[Fuente et al.(1995)]{1995ApJ...442L..33F} Fuente, A., 
Martin-Pintado, J., \& Gaume, R.\ 1995, \apjl, 442, L33 

\bibitem[Fuente et 
al.(1996)]{1996A&A...312..599F} Fuente, A., Rodr\'{\i}guez-Franco, A., \& Mart\'{\i}n-Pintado, J.\ 1996, \aap, 312, 599 

\bibitem[Fuente et 
al.(2003)]{2003A&A...406..899F} Fuente, A., Rodr{\i}guez-Franco, A., Garc{\i}a-Burillo, S., Mart{\i}n-Pintado, J., \& Black, J.~H.\ 2003, \aap, 406, 899 

\bibitem[Fuente et al.(2005)]{2005ESASP.577..281F} Fuente, A., 
Garc{\'{\i}}a-Burillo, S., Gerin, M., et al.\ 2005, ESA Special 
Publication, 577, 281 

\bibitem[Fuente et al.(2006)]{2006ApJ...641L.105F} Fuente, A., 
Garc{\'{\i}}a-Burillo, S., Gerin, M., et al.\ 2006, \apjl, 641, L105 

\bibitem[Fuente et 
al.(2008)]{2008A&A...492..675F} Fuente, A., Garc{\'{\i}}a-Burillo, S., Usero, A., et al.\ 2008, \aap, 492, 675 

\bibitem[Garc{\'{\i}}a-Burillo et 
al.(2000)]{2000A&A...355..499G} Garc{\'{\i}}a-Burillo, S., Mart{\'{\i}}n-Pintado, J., Fuente, A., \& Neri, R.\ 2000, \aap, 355, 499 

\bibitem[Garc{\'{\i}}a-Burillo et al.(2001)]{2001ApJ...563L..27G} 
Garc{\'{\i}}a-Burillo, S., Mart{\'{\i}}n-Pintado, J., Fuente, A., 
\& Neri, R.\ 2001, \apjl, 563, L27 

\bibitem[Garc{\'{\i}}a-Burillo et al.(2002)]{2002ApJ...575L..55G} 
Garc{\'{\i}}a-Burillo, S., Mart{\'{\i}}n-Pintado, J., Fuente, A., Usero, 
A., \& Neri, R.\ 2002, \apjl, 575, L55 

\bibitem[Goicoechea et 
al.(2006)]{2006A&A...456..565G} Goicoechea, J.~R., Pety, J., Gerin, M., et al.\ 2006, \aap, 456, 565 

\bibitem[Gre02]{2002A&A...383...56G} Greve, A., Wills, K.~A., Neininger, N., \& Pedlar, A.\ 2002, \aap, 383, 56 

\bibitem[Greve(2011)]{2011A&A...529A..51G} Greve, A.\ 2011, \aap, 529, AA51 

\bibitem[Guzm{\'a}n et al.(2014)]{2014arXiv1404.7798G} Guzm{\'a}n, V.~V., 
Pety, J., Gratier, P., et al.\ 2014, arXiv:1404.7798 

\bibitem[Guzm{\'a}n et 
al.(2013)]{2013A&A...560A..73G} Guzm{\'a}n, V.~V., Goicoechea, J.~R., Pety, J., et al.\ 2013, \aap, 560, AA73 

\bibitem[Indriolo et al.(2014)]{2014arXiv1412.1106I} Indriolo, N., Neufeld, 
D.~A., Gerin, M., et al.\ 2014, arXiv:1412.1106 

\bibitem[Joy et al.(1987)]{1987ApJ...319..314J} Joy, M., Lester, D.~F., 
\& Harvey, P.~M.\ 1987, \apj, 319, 314 

\bibitem[Kronberg et al.(1981)]{1981ApJ...246..751K} Kronberg, P.~P., 
Biermann, P., \& Schwab, F.~R.\ 1981, \apj, 246, 751 

\bibitem[Le Petit et al.(2006)]{2006ApJS..164..506L} Le Petit, F., 
Nehm{\'e}, C., Le Bourlot, J., \& Roueff, E.\ 2006, \apjs, 164, 506 

\bibitem[Lester et al.(1990)]{1990ApJ...352..544L} Lester, D.~F., Gaffney, 
N., Carr, J.~S., \& Joy, M.\ 1990, \apj, 352, 544 

\bibitem[Li et al.(2015)]{2015ApJS..216....6L} Li, S., de Grijs, R., 
Anders, P., \& Li, C.\ 2015, \apjs, 216, 6 

\bibitem[Liszt 
\& Lucas(2001)]{2001A&A...370..576L} Liszt, H., \& Lucas, R.\ 2001, \aap, 370, 576 

\bibitem[Mao et 
al.(2000)]{2000A&A...358..433M} Mao, R.~Q., Henkel, C., Schulz, A., et al.\ 2000, \aap, 358, 433 

\bibitem[Mart{\'{\i}}n et 
al.(2006)]{2006A&A...450L..13M} Mart{\'{\i}}n, S., Mart{\'{\i}}n-Pintado, J., \& Mauersberger, R.\ 2006, \aap, 450, L13 

\bibitem[Martin-Pintado et 
al.(1992)]{1992A&A...254..315M} Mart\'{\i}n-Pintado, J., Bachiller, R., \& Fuente, A.\ 1992, \aap, 254, 315 

\bibitem[Murray(2011)]{2011ApJ...729..133M} Murray, N.\ 2011, \apj, 729, 
133 

\bibitem[Nei98]{1998A&A...339..737N} Neininger, N., Guelin, M., Klein, U., Garcia-Burillo, S., \& Wielebinski, R.\ 1998, \aap, 339, 737 

\bibitem[Origlia et al.(2004)]{2004ApJ...606..862O} Origlia, L., Ranalli, 
P., Comastri, A., \& Maiolino, R.\ 2004, \apj, 606, 862 

\bibitem[Pety et 
al.(2005)]{2005A&A...435..885P} Pety, J., Teyssier, D., Foss{\'e}, D., et al.\ 2005, \aap, 435, 885 

\bibitem[Pilleri et 
al.(2013)]{2013A&A...554A..87P} Pilleri, P., Trevi{\~n}o-Morales, S., Fuente, A., et al.\ 2013, \aap, 554, AA87 

\bibitem[Salas et al.(2014)]{2014ApJ...797..134S} Salas, P., Galaz, G., 
Salter, D., et al.\ 2014, \apj, 797, 134 

\bibitem[She95]{1995ApJ...445L..99S} Shen, J., \& Lo, K.~Y.\ 1995, \apjl, 445, L99 


\bibitem[Shopbell 
\& Bland-Hawthorn(1998)]{1998ApJ...493..129S} Shopbell, P.~L., \& Bland-Hawthorn, J.\ 1998, \apj, 493, 129 

\bibitem[Spaans 
\& Meijerink(2007)]{2007ApJ...664L..23S} Spaans, M., \& Meijerink, R.\ 2007, \apjl, 664, L23 

\bibitem[Strickland 
\& Heckman(2007)]{2007ApJ...658..258S} Strickland, D.~K., \& Heckman, T.~M.\ 2007, \apj, 658, 258 

\bibitem[Suchkov et al.(1993)]{1993ApJ...413..542S} Suchkov, A., Allen, 
R.~J., \& Heckman, T.~M.\ 1993, \apj, 413, 542 

\bibitem[Umeda et al.(2002)]{2002ApJ...578..855U} Umeda, H., Nomoto, K., 
Tsuru, T.~G., \& Matsumoto, H.\ 2002, \apj, 578, 855 

\bibitem[Usero et 
al.(2006)]{2006A&A...448..457U} Usero, A., Garc{\'{\i}}a-Burillo, S., Mart{\'{\i}}n-Pintado, J., Fuente, A., \& Neri, R.\ 2006, \aap, 448, 457 

\bibitem[Viti et 
al.(2014)]{2014A&A...570A..28V} Viti, S., Garc{\'{\i}}a-Burillo, S., Fuente, A., et al.\ 2014, \aap, 570, AA28 

\bibitem[Walter et al.(2002)]{2002ApJ...580L..21W} Walter, F., Weiss, A., 
\& Scoville, N.\ 2002, \apjl, 580, L21 

\bibitem[Wei{\ss} et 
al.(1999)]{1999A&A...345L..23W} Wei{\ss}, A., Walter, F., Neininger, N., \& Klein, U.\ 1999, \aap, 345, L23 

\bibitem[Wei{\ss} et al.(2001a)]{2001ApJ...554L.143W} Wei{\ss}, A., 
Neininger, N., Henkel, C., Stutzki, J., 
\& Klein, U.\ 2001a, \apjl, 554, L143 

\bibitem[Wei{\ss} et 
al.(2001b)]{2001A&A...365..571W} Wei{\ss}, A., Neininger, N., H{\"u}ttemeister, S., \& Klein, U.\ 2001b, \aap, 365, 571 

\bibitem[Wills et al.(1999)]{1999MNRAS.309..395W} Wills, K.~A., Redman, 
M.~P., Muxlow, T.~W.~B., \& Pedlar, A.\ 1999, \mnras, 309, 395 

\bibitem[Wil00]{2000MNRAS.316...33W} Wills, K.~A., Das, M., 
Pedlar, A., Muxlow, T.~W.~B., \& Robinson, T.~G.\ 2000, \mnras, 316, 33 




\end{thebibliography}

\newpage
\begin{appendix}
\renewcommand\thefigure{\thesection.\arabic{figure}}    
\section{Tables and Figures}
\setcounter{figure}{0}  

\begin{figure*} 
 \centering
 \includegraphics[width=0.9\textwidth] {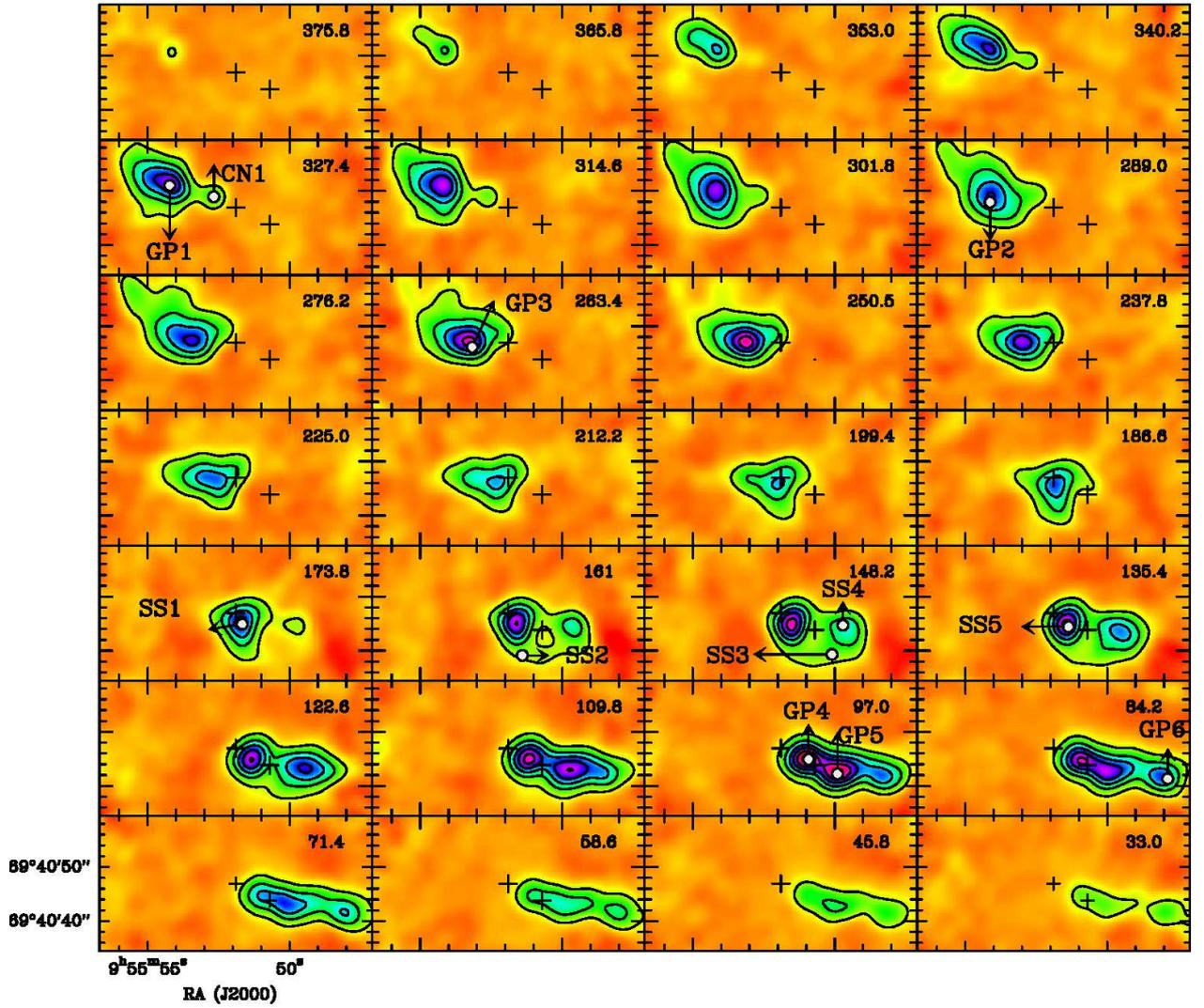}
 \caption{Spectral maps of the CN 1$\rightarrow$0 line convolved to an angular resolution of 3.8$\arcsec$. To increase the S/N ratio, the velocity resolution 
has been degraded to 12.8~km~s$^{-1}$. The number in the right-upper corner indicates the central channel velocity. Contour levels are 
0.17 (5$\times$$\sigma$) to 1.955 in steps of 0.255~K. Crosses indicate the dynamical center of the galaxy and 
the position of SN~41.9+58.}
 \label{figa1}
\end{figure*}

\begin{figure*} 
 \centering
 \includegraphics[width=0.9\textwidth] {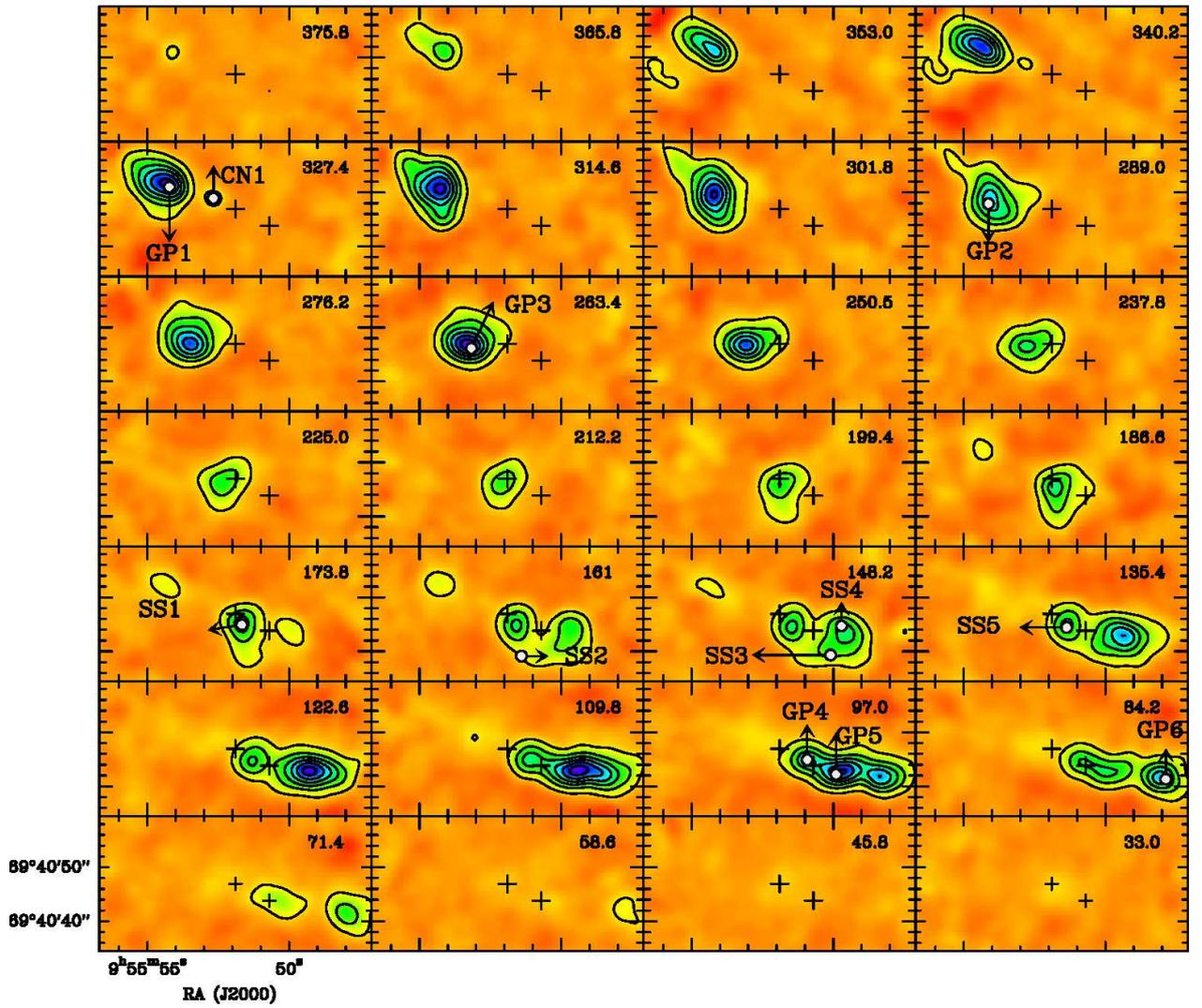}
 \caption{The same as Fig.~A.1 for the CS 3$\rightarrow$2 line. Contour levels are 0.065 (3$\times$$\sigma$) to 1.495 in 
steps of 0.13~K. The eastern knot at velocities $\sim$144-192~km~s$^{-1}$ is due to the emission of the H(35)$\alpha$ line.}
 \label{figa2}
\end{figure*}

\begin{figure*} 
 \centering
 \includegraphics[width=0.9\textwidth] {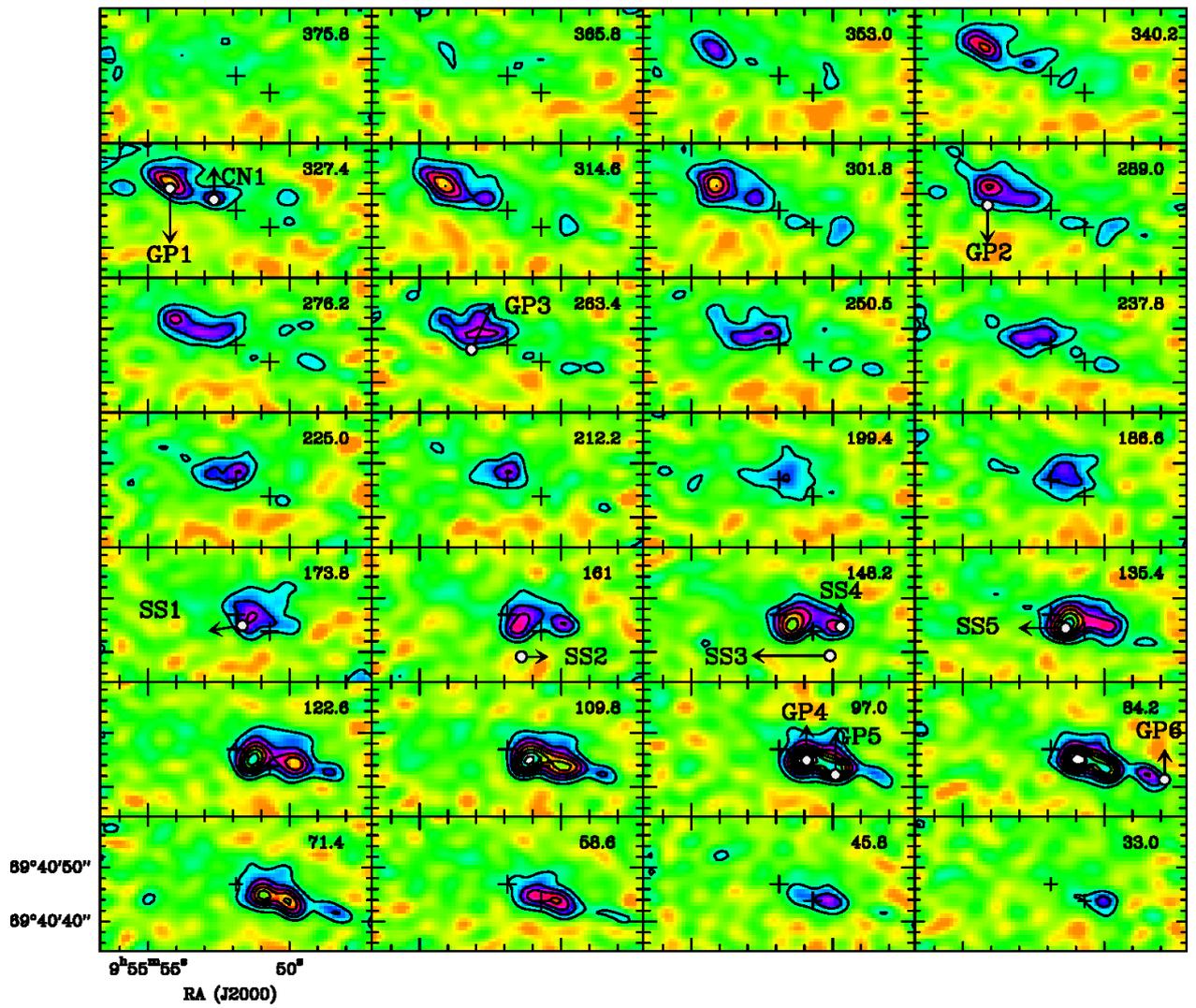}
 \caption{The same as Fig.~A.1 for the H(41)$\alpha$ line. First contour and step are 0.028~K ($\sim$3$\times$$\sigma$).}
 \label{figa3}
\end{figure*}

\begin{table*}
\caption{Gaussians parameters to the observed spectra$^1$}
\begin{center}    
\begin{tabular}{p{1.1cm} p{1.6cm} p{1.4cm} p{2.4cm} p{2.4cm} p{2.4cm} p{1.8cm} p{0cm}} 
\\
\hline
\multicolumn{7}{c}{E1 (+14$\arcsec$, +5$\arcsec$)} & \\
\multicolumn{2}{l}{Line} & \multicolumn{1}{c}{Freq(GHz)}   & 
\multicolumn{1}{c}{I(K km~s$^{-1}$)} & \multicolumn{1}{c}{v$_{lsr}$(km~s$^{-1}$)} & 
\multicolumn{1}{c}{$\Delta$v(km~s$^{-1}$)} & \multicolumn{1}{c}{T$_{MB}$(K)} \\ 
\hline
CN              &  1$\rightarrow$0           &  113.491   & 64.15 (0.56)  &  316.26 (0.33)  &  75.73 (0.77)  & 0.79  \\  
H$^{13}$CO$^+$  &  1$\rightarrow$0           &  86.754    & 4.52 (0.67)   &  324.66 (5.02)  &  69.39 (12.4) & 0.06  \\ 
H               & 41$\alpha$                 &  92.034    & 5.42 (0.19)   &  320.48 (1.21)  &  70.79 (2.89)  & 0.07  \\
CH$_3$CN        &  5$_k$$\rightarrow$4$_k$$^*$   & 91.987 & 2.43 (0.20)   &  328.58 (3.13)  &  74.26 (6.62)  & 0.03  \\
N$_2$H$^+$      &  1$\rightarrow$0           & 93.173     & 6.98 (0.17)  &  321.42 (0.73)   & 60.23 (1.57) & 0.11  \\ 
C$^{18}$O       &  1$\rightarrow$0           & 109.782    & 22.74 (0.62)  &  324.40 (0.93)  &  69.58 (2.18)  &  0.31  \\
C$_3$H$_2$     &  3$\rightarrow$2            & 145.089    &  9.90 (0.24)  &  315.23 (0.75)  &  63.77 (1.69)  &  0.14 \\
H$_2$CO        &  2$_{0,2}$$\rightarrow$1$_{0,1}$ & 145.603    &  8.45 (0.20) &  321.30 (0.67) & 56.90 (1.58) &  0.14  \\
HC$_3$N        &  16$\rightarrow$15               & 145.561    &  4.35 (0.17) &  330.32 (0.90) & 43.12 (1.96) &  0.09 \\
CS             &  3$\rightarrow$2           & 145.760     & 31.84 (0.37) &  322.56 (0.32)  &  55.63 (0.71) &  0.54 \\
\hline
\multicolumn{7}{c}{E2 (+5$\arcsec$, +2$\arcsec$)} & \\
\multicolumn{2}{l}{Line} & \multicolumn{1}{c}{Freq(GHz)}   & 
\multicolumn{1}{c}{I(K km~s$^{-1}$)} & \multicolumn{1}{c}{v$_{lsr}$(km~s$^{-1}$)} & 
\multicolumn{1}{c}{$\Delta$v(km~s$^{-1}$)} & \multicolumn{1}{c}{T$_{MB}$(K)} \\ 
\hline
CN             & 1$\rightarrow$0   &  113.491   &  56.09 (0.53) &  249.23 (0.35) &  79.56 (0.92) &  0.66 \\
               &                   &            &   4.83 (0.30) &  330.76 (0.78) &  26.26 (1.70) &  0.17\\
H$^{13}$CO$^+$ &  1$\rightarrow$0  &   86.754   &   2.19 (0.30) &  258.66 (4.28) &  57.14 (7.48)   &  0.03  \\
H              &  41$\alpha$       &   92.034   &   7.25 (0.24) &  278.04 (1.89) &  121.01 (5.46)  &  0.06 \\
CH$_3$CN       &  5$_k$$\rightarrow$4$_k$$^*$   & 91.987 &   1.35 (0.20) &  284.18 (5.79) &  85.36 (13.7) & 0.01     \\ 
N$_2$H$^+$     &  1$\rightarrow$0  &   93.173   &  3.74 (0.15)  &  255.80 (1.08)  &  58.24 (2.89)  & 0.06 \\  
C$^{18}$O      &  1$\rightarrow$0  &  109.782   &  9.22 (0.65)  &  259.28 (1.78)  &  53.14 (4.66)  & 0.16 \\
C$_3$H$_2$     &  3$\rightarrow$2  &  145.089   &  4.95 (0.17) &  245.88 (0.88) &   53.58 (2.13) &  0.09 \\               
               &                   &            &  0.57 (0.12) &  311.92 (2.90) &   26.65 (5.71) &  0.02 \\
H$_2$CO       &  2$_{0,2}$$\rightarrow$1$_{0,1}$ & 145.603  &  4.34 (0.16) &  256.14 (0.77) &   45.57 (1.98) &  0.09  \\
HC$_3$N+H$_2$CO$^{2}$   &                       & 145.561  &  2.85 (0.17) &  253.33 (1.48) &   53.47 (3.90) &  0.05 \\
CS             &  3$\rightarrow$2                 & 146.969  &  16.83 (0.15) &  258.63 (0.20)  &  46.79 (0.53) &  0.34 \\
               &                                  &          &   1.43 (0.11) &  329.47 (0.94)  &  25.13 (2.28) &  0.05  \\
\hline
\multicolumn{7}{c}{W2 (-5$\arcsec$, -2$\arcsec$)} & \\
\multicolumn{2}{l}{Line} & \multicolumn{1}{c}{Freq(GHz)}   &
\multicolumn{1}{c}{I(K km~s$^{-1}$)} & \multicolumn{1}{c}{v$_{lsr}$(km~s$^{-1}$)} & 
\multicolumn{1}{c}{$\Delta$v(km~s$^{-1}$)} & \multicolumn{1}{c}{T$_{MB}$(K)} \\ 
\hline 
CN & 1$\rightarrow$0               &  113.491   &  10.65 (0.82) &   95.78 (0.37) &   25.788 (1.18) & 0.39 \\
   &                                &            &  56.67 (0.59) &   99.31 (0.62) &   92.166 (2.18) & 0.58 \\
   &                                &            &  20.75 (0.54) &  156.73 (1.09) &   68.352 (1.92) & 0.28 \\
H$^{13}$CO$^+$ &  1$\rightarrow$0  &   86.754   &   2.86 (0.50) &  115.70 (10.6) &  111.91 (18.2) & 0.02 \\ 
H              &  41$\alpha$       &   92.034   &  10.75 (0.29) &  103.65 (0.59) &   81.50 (1.71) & 0.12 \\
               &                    &            &  8.87  (0.39) &  172.86 (7.51) &  352.24 (18.08) & 0.02 \\
CH$_3$CN       &  5$_k$$\rightarrow$4$_k$$^*$   &  91.987       &   &            &                  &    $<$0.015 \\
N$_2$H$^+$     &  1$\rightarrow$0  & 93.173     &  3.85 (0.13)  &   93.13 (6.28) &   48.56 (6.28)   &  0.07 \\
               &                   &            &  1.47 (0.13)  &  145.67 (6.28) &   35.74 (6.28)   &  0.04 \\
               &                   &            &  0.91 (0.13)  &  183.60 (6.28) &   30.43 (6.28)   &  0.03 \\
C$^{18}$O      &  1$\rightarrow$0  & 109.782    & 10.45 (0.60)  &  130.29 (3.30) & 111.32 (7.20)  & 0.09 \\
C$_3$H$_2$     &  3$\rightarrow$2  & 145.089    &  4.72 (0.20)  &  89.20 (1.04)  &  55.84 (2.74) & 0.08 \\
               &                   &            &  2.98 (0.21)  & 162.01 (0.31)  &  64.91 (4.64) & 0.04 \\
H$_2$CO$^*$    &  2$_{0,2}$$\rightarrow$1$_{0,1}$  & 145.603   &  2.96 (0.25)  &  95.22 (0.90) &   37.15 (2.58) &  0.07 \\
HC$_3$N+H$_2$CO$^3$    &        & 145.561   &  8.20 (0.70)  &  93.10 (3.30) &   93.54 (10.92) &  0.08 \\
CS             &  3$\rightarrow$2                  & 146.969   &  10.96 (2.52)  &   96.54 (3.24)  &  37.31 (4.49) &  0.27 \\
               &                                   &           &  7.77  (3.55)  &   136.99 (4.27) &  42.03 (16.7) &  0.17 \\
               &                                   &           &  4.22  (1.47)  &   182.91 (4.92) &  34.698.89) &  0.11 \\
\hline
\multicolumn{7}{c}{W1 (-14$\arcsec$, -5$\arcsec$)} & \\ 
\multicolumn{2}{l}{Line} & \multicolumn{1}{c}{Freq(GHz)}   & 
\multicolumn{1}{c}{I(K km~s$^{-1}$)} & \multicolumn{1}{c}{v$_{lsr}$(km~s$^{-1}$)} & 
\multicolumn{1}{c}{$\Delta$v(km~s$^{-1}$)} & \multicolumn{1}{c}{T$_{MB}$(K)} \\ 
\hline
CN             & 1$\rightarrow$0  &  113.491   & 62.38 (0.58) &  103.05 (0.36)  &  80.399 (0.87)  &  0.73  \\ 
H$^{13}$CO$^+$ &  1$\rightarrow$0 &   86.754   &  3.90 (0.57) &  104.39 (5.51)  &  76.888 (13.54) &  0.05  \\  
H              &  41$\alpha$                   &  92.034   &  4.76 (0.18)  &  98.51 (1.52)   &  80.41 (3.65) &  0.06    \\ 
CH$_3$CN       &  5$_k$$\rightarrow$4$_k$$^*$  &  91.987   &  1.78 (0.17)  &  127.93 (3.60)  &  71.73 (7.26) &  0.02  \\   
N$_2$H$^+$     &  1$\rightarrow$0 &   93.173   &  6.61 (0.16)  &  113.57 (0.74)  &  62.23 (1.81)  &  0.10 \\  
C$^{18}$O      &  1$\rightarrow$0 &  109.782 & 24.90 (0.51) &  118.46 (0.65)  &  63.32 (1.45) &  0.37 \\
C$_3$H$_2$     &  3$\rightarrow$2 &  145.089 & 13.00 (0.72) &  103.29 (1.55)  &  57.80 (3.92) &  0.21  \\
H$_2$CO        &  2$_{0,2}$$\rightarrow$1$_{0,1}$ & 145.603 &  6.12 (0.51) &  110.67 (1.18)  &  47.71 (2.57) & 0.12 \\
HC$_3$N        &  16$\rightarrow$15               & 145.561 &  3.66 (1.52) &  100.41 (18.4)  &  93.16 (43.1) & 0.04 \\
CS             &  3$\rightarrow$2                 & 145.760 &  29.52 (0.26)   &  116.01  (0.24) &  54.22 (0.55)  &  0.51  \\
\hline
\\
\end{tabular}
\end{center}

\noindent
$^1$The PdBI images have been convolved to the same angular resolution (5.9$\arcsec$ x 5.9$\arcsec$).
$^*$The k=0,1,2 and 3 components are blended. The velocity of the Gaussian fit is calculated relative to the frequency of the k=0 component, $\nu$=91987~MHz.
$^{2}$The $\sim$326~km~s$^{-1}$ component of H$_2$CO overlap with the $\sim$270~km~s$^{-1}$ component of HC$_3$N.
$^{3}$A possible $\sim$180~km~s$^{-1}$ component of H$_2$CO would overlap with the $\sim$93~km~s$^{-1}$ component of HC$_3$N.

\end{table*}

\end{appendix}

\end{document}